\documentclass[12pt,preprint]{aastex}
\usepackage{amsmath,euscript,amsfonts}

\def\simlt{\mathrel{\hbox{\rlap{\hbox{\lower4pt\hbox{$\sim$}}}\hbox{$<$}}}}
\def\simgt{\mathrel{\hbox{\rlap{\hbox{\lower4pt\hbox{$\sim$}}}\hbox{$>$}}}}

\def\ale{\mathrel{\hbox{\rlap{\hbox{\lower4pt\hbox{$\sim$}}}\hbox{$<$}}}}
\def\age{\mathrel{\hbox{\rlap{\hbox{\lower4pt\hbox{$\sim$}}}\hbox{$>$}}}}

\def\nodata{---}

\def\cit{1}
\def\uva{2}
\def\vla{3}

\shorttitle{The Radio Luminous SN\,2003bg}
\shortauthors{Soderberg et al.}

\begin{document}

\title{The Radio and X-ray Luminous SN\,2003bg and the Circumstellar
  Density Variations Around Radio Supernovae}

\author{
A.~M. Soderberg\altaffilmark{\cit}, 
R.~A. Chevalier\altaffilmark{\uva},
S.~R. Kulkarni\altaffilmark{\cit}, 
D.~A. Frail\altaffilmark{\vla},
}

\altaffiltext{\cit}{Division of Physics, Mathematics and Astronomy,
        105-24, California Institute of Technology, Pasadena, CA
        91125}
\altaffiltext{\uva}{Department of Astronomy, University of Virginia, P.O. Box 3818, Charlottesville, VA 22903-0818}
\altaffiltext{\vla}{National Radio Astronomy Observatory, Socorro, NM 87801}

\begin{abstract}
We report extensive radio and X-ray observations of SN\,2003bg whose
spectroscopic evolution shows a transition from a broad-lined Type Ic
to a hydrogen-rich Type II and later to a typical hydrogen-poor Type
Ibc.  We show that the extraordinarily luminous radio emission is well
described by a self-absorption dominated synchrotron spectrum while
the observed X-ray emission at $t\approx 30$ days is adequately fit by
Inverse Compton scattering of the optical photons off of the
synchrotron emitting electrons.  Our radio model implies a
sub-relativistic ejecta velocity, $\overline{v}\approx 0.24 c$, at
$t_0\approx 10$ days after the explosion which emphasizes that broad
optical absorption lines do not imply relativistic ejecta.  We find
that the total energy of the radio emitting region evolves as
$E\approx 7.3\times 10^{48} (t/t_0)^{0.4}$ erg assuming equipartition
of energy between relativistic electrons and magnetic fields
($\epsilon_e=\epsilon_B=0.1$).  The circumstellar density is well
described by a stellar wind profile with modest (factor of $\sim 2$)
episodic density enhancements which produce abrupt achromatic flux
variations.  We estimate an average mass loss rate of $\dot{M}\approx
3 \times 10^{-4}~\rm M_{\odot}~yr^{-1}$ (assuming a wind velocity of
$v_w=10^3~\rm km~s^{-1}$) for the progenitor, consistent with the
observed values for Galactic Wolf-Rayet stars.  Comparison with other
events reveals that $\sim 50\%$ of radio supernovae show similar
short timescale flux variations attributable to circumstellar density
irregularities.  Specifically, the radio light-curves of SN\,2003bg
are strikingly similar to those of the Type IIb SN\,2001ig, suggestive
of a common progenitor evolution for these two events.  Based on the
relative intensity of the inferred density enhancements, we conclude
that the progenitors of SNe 2003bg and 2001ig experienced
quasi-periodic mass loss episodes just prior to the SN explosion.
Finally, this study emphasizes that abrupt radio light-curve
variations cannot be used as a reliable proxy for an engine-driven
explosion, including off-axis gamma-ray bursts.
\end{abstract}


\section{Introduction}
\label{sec:intro}

Accounting for $\sim 10\%$ of the nearby supernova population, Type
Ibc supernovae (hereafter SNe Ibc) are identified by their lack of
spectroscopic hydrogen and silicon features (see \citealt{f97} for a
review).  Recognized as a rare subclass of core-collapse supernova
$\sim 20$ years ago \citep{emn+85}, SNe Ibc have recently enjoyed a
revitalized interest. Beginning with the discovery of Type Ic
SN\,1998bw in temporal and spatial coincidence with gamma-ray burst
(GRB)\,980425 \citep{gvv+98,paa+00}, we now know that most
long-duration GRBs (e.g. \citealt{smg+03}) and X-ray flashes (XRFs;
\citealt{skf+05}) are associated with SNe Ibc. However, radio
observations of local SNe Ibc indicate that the inverse is {\em not}
true; there is a strict limit of $\lesssim 10\%$ on the fraction of
SNe Ibc that could be accompanied by a GRB or XRF \citep{snk+05}.  The
lack of hydrogen features in both SNe Ibc and GRB/XRF-associated SNe
imply that they represent the explosion of massive stripped-core
progenitors \citep{wl85,wes99}.

The popular models for SNe Ibc progenitors include massive Wolf-Rayet
(WR) stars that eject their envelopes through strong dense winds, and
close binary systems where the progenitor star is stripped of its
hydrogen-rich layer by the companion \citep{ew88}. Despite dedicated
archival searches, the progenitors of SNe Ibc are still poorly
constrained by pre-explosion images.  The best photometric constraints
are currently associated with SN\,2002ap and SN\,2004gt which exclude
WR progenitors in the top $\sim 30\%$ and $\sim 50\%$ of the
population, respectively \citep{svr+02,gfk+05,mss05}.  While
intriguing, these archival observations lack the sensitivity to
clearly discriminate between the WR and binary progenitor models.

By studying the circumstellar medium around the supernova, however, it
is possible to place independent constraints on the progenitor.
Observations show that Galactic Wolf-Rayet stars are embedded in wind
stratified media \citep{chu02} while binary systems are associated
with disrupted circumstellar media, possibly including an outflow during a
common envelope phase \citep{pjh92}.

Radio observations provide the most direct probe of the circumstellar
density structure around supernovae \citep{wsp+86,c98}.
By observing the dynamical interaction of the ejecta with the
surrounding medium we are able to map out the mass loss history of the
progenitor star.  The identification of irregular density profiles may
therefore distinguish between Wolf-Rayet and binary progenitor systems.

Here we present extensive radio observations of SN\,2003bg, discovered
as part of our ongoing radio survey of local Type Ibc supernovae.  The
peculiar SN\,2003bg spectroscopically evolved from a broad-lined Type
Ic to a hydrogen-rich Type II and later to a typical Type Ibc (Hamuy
{\it et al.}, in prep), thereby bridging the hydrogen-rich and poor
divisions of the core-collapse classification system. Our
densely-sampled radio light-curves show that the extraordinarily
luminous radio emission for a SN Ibc is characterized by episodic
short-timescale variations. We show that these variations are well
described by abrupt density enhancements in the circumstellar medium.
Comparison with other radio supernovae shows that $\sim 50\%$ of all
well-studied events similarly show evidence for abrupt light-curve
variations. We review the mass loss evolution observed (and inferred)
for the progenitors of these core-collapse SNe.  Using the observed
radio properties for the peculiar SN\,2003bg, we place constraints on
the nature of its progenitor system.

The organization of this paper is as follows: observations from the
Very Large Array (VLA) and the {\it Chandra} X-ray Observatory (CXO)
are described in \S\ref{sec:obs}.  Preliminary estimates of the
energy, velocity and density of the radio emitting region are presented
in \S\ref{sec:ep}.  Modeling of the radio light-curves are presented
in \S\ref{sec:model} and \S\ref{sec:ssa} while a discussion of the
radio polarization mechanism follows as \S\ref{sec:pol_disc}.  Our
modeling of the X-ray emission is discussed in \S\ref{sec:xray_model}.
In \S\ref{sec:bumps} we present a compilation of radio supernovae with
abrupt light-curve variations and review their circumstellar
irregularities.  Finally, in \S\ref{sec:prog} we discuss the possible
causes of the density enhancements surrounding SN\,2003bg and the
implications for the progenitor system.

\section{Observations}
\label{sec:obs}

SN\,2003bg was optically discovered on 2003 February 25.7 UT, offset
16''.3 W and 24".6 S from the center of host galaxy, MCG -05-10-15, at
$d\approx 19.6$ Mpc \citep{wc03}.  Early spectroscopy on 2003 February
28 UT indicated that SN\,2003bg was a peculiar Type Ic supernova with
broad optical absorption lines, indicative of fast photospheric
velocities and comparable to those seen in SN\,1998bw \citep{fc03}.
Soon thereafter, SN\,2003bg developed strong, broad H$\alpha$ emission
with a P-Cygni absorption component indicating an expansion velocity
of $-17,110~\rm km~s^{-1}$ \citep{hpt03}.  P-Cygni components were
additionally observed for H$\beta$ and H$\gamma$, prompting the
reclassification of SN\,2003bg as an unusual Type II supernova.  Nebular
spectra taken several months later revealed that all evidence of
hydrogen had disappeared, suggestive that the SN had
transitioned back to a Type Ibc event (Hamuy et al., in prep).  

The peak time of the optical light-curves suggests that SN\,2003bg was
discovered within just a few days of explosion (Hamuy, private
communication).  Throughout this paper we therefore assume an
approximate explosion date of 2003 February 22 UT.

\subsection{Very Large Array Data}
\label{sec:obs_vla}
\begin{figure}
\plotone{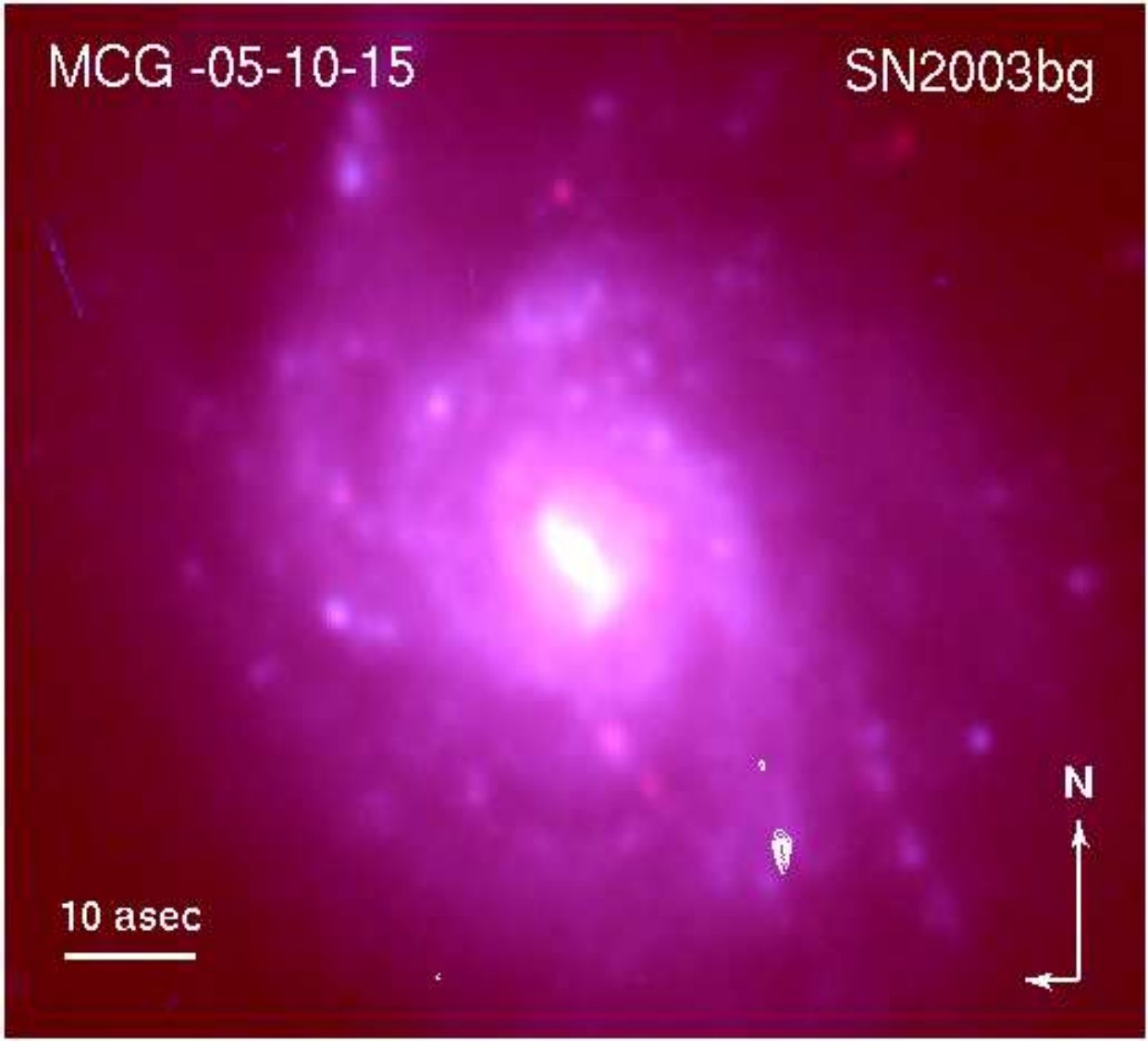}
\caption{Composite color ($g'r'$) image of host galaxy MCG -05-10-15
  taken with the Palomar 200-inch telescope with the Large Field
  Camera.  White contours map the radio emission from SN\,2003bg
  detected with the Very Large Array at 8.46 GHz on 2005 May 31 (VLA
  B-array configuration).  Contours are linearly spaced and
  correspond to flux density values between 0.5 and 4.0 mJy.  The SN
  was located in a star-forming spiral arm of the host galaxy at
  position $\alpha\rm (J2000)=04^{\rm h}10^{\rm m}59.42^{\rm s},
  \delta\rm (J2000)=-31^{\rm o}24'50.3''$ ($\pm$0.1 arcsec in each
  coordinate).  We note that no diffuse emission from the host galaxy
  was detected in any of our radio observations.
\label{fig:rgb}}
\end{figure}

Motivated by the spectroscopic similarity of SN\,2003bg to the radio
luminous SN\,1998bw, we initiated radio observations of SN\,2003bg on
2003 March 4.07 UT with the Very Large Array\footnote{The Very Large
Array and Very Long Baseline Array are operated by the National Radio
Astronomy Observatory, a facility of the National Science Foundation
operated under cooperative agreement by Associated Universities, Inc.}
(VLA).  At 8.46 GHz we detected a radio source coincident with the
optical position at $\alpha\rm (J2000)=04^{\rm h}10^{\rm m}59.42^{\rm
s}, \delta\rm (J2000)=-31^{\rm o}24'50.3''$ ($\pm$0.1 arcsec in each
coordinate) with flux density of $f_{\nu}=2.51\pm 0.05$ mJy
(Figure~\ref{fig:rgb}).  We subsequently began an intense follow-up
campaign to study the temporal and spectral evolution of the radio
emission.

Radio data were collected at 1.43, 4.86, 8.46, 15.0, 22.5 and 43.3 GHz
spanning 2003 March through 2005 October and are summarized in
Table~\ref{tab:vla}.  All VLA observations were taken in standard
continuum observing mode with a bandwidth of $2\times 50$ MHz.  At
22.5 and 43.3 GHz we included referenced pointing scans to correct for
the systematic 10-20 arcsec pointing errors of the VLA antennas.  We
used 3C48, 3C147 and 3C286 (J0137+331, J0542+498 and J1331+305) for
flux calibration, while phase referencing was performed against
calibrators J0407-330 and J0453-281.  Data were reduced using standard
packages within the Astronomical Image Processing System (AIPS).  No
diffuse radio emission from the host galaxy was detected in any of our
observations.  Flux density measurements were obtained by fitting a
Gaussian model to the SN.  In addition to the {\it rms} noise in each
measurement, we include a systematic uncertainty of $2\%$ due to the
uncertainty in the absolute flux calibration.

\begin{figure}
\plotone{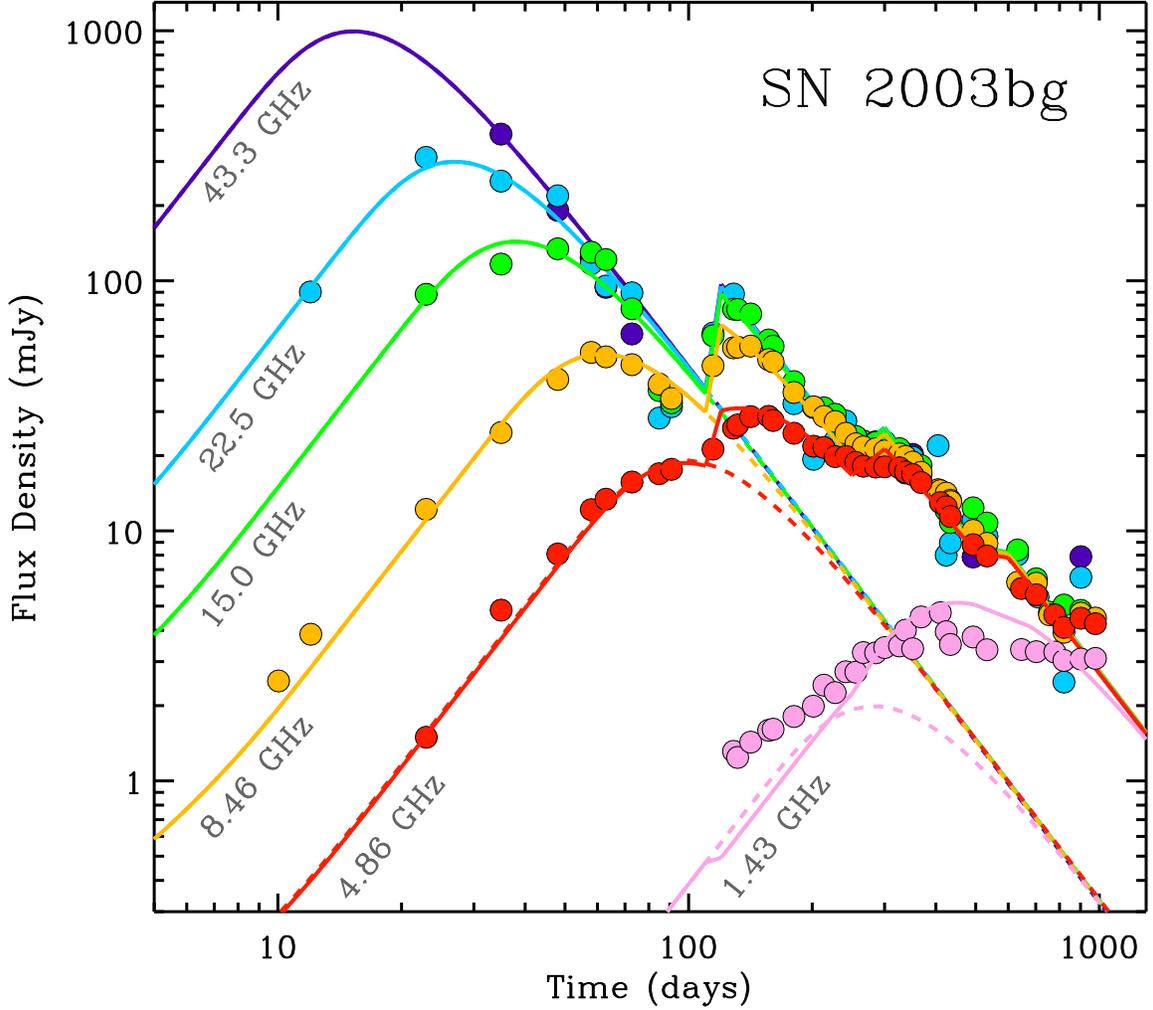}
\caption{Radio light-curves of SN\,2003bg. From bottom to top, the
  measurements were taken with the VLA at 1.43 GHz (pink), 4.86 GHz
  (red), 8.46 GHz (orange), 15.0 GHz (green), 22.5 GHz (blue) and 43.3
  GHz (purple) between 2003 March and 2005 October UT ($t\sim 10-1000$
  days after the explosion).  The data have been scaled to the 8.46
  GHz measurements by $\nu^{\beta}$ where $\beta\approx -1.1$ is the
  observed optically thin spectral index.  An abrupt, achromatic
  light-curve deviation is observed at $t\sim 120$ days and
  additional, weaker deviations are suggested at $t\sim 300$ and 600
  days.  The synchrotron self-absorption model described in
  \S\ref{sec:ssa} is overplotted (solid lines).  An extrapolation of
  the early fit ($t\lesssim 110$ days) is shown for comparison (dashed
  lines).
\label{fig:oneplot}}
\end{figure}

The SN\,2003bg radio light-curves are compiled in
Figures~\ref{fig:oneplot} and \ref{fig:lt_curves}, spanning $\sim 10$
to $\sim 1000$ days after the explosion.  The radio evolution of
SN\,2003bg is characterized by several achromatic short timescale
($\delta t/t \lesssim 1$) variations.  As will be discussed in
\S\ref{sec:bumps}, light-curve variations have also been observed for
$\sim 50\%$ of well-sampled radio supernovae including Type IIL
SN\,1979C \citep{wvd+91}, Type Ic SN\,1998bw \citep{kfw+98,wpm01}, and
more recently for Type IIb SN\,2001ig \citep{rss+04}.

\subsubsection{Radio Spectrum}
\label{sec:radio_spec}

\begin{figure}
\plotone{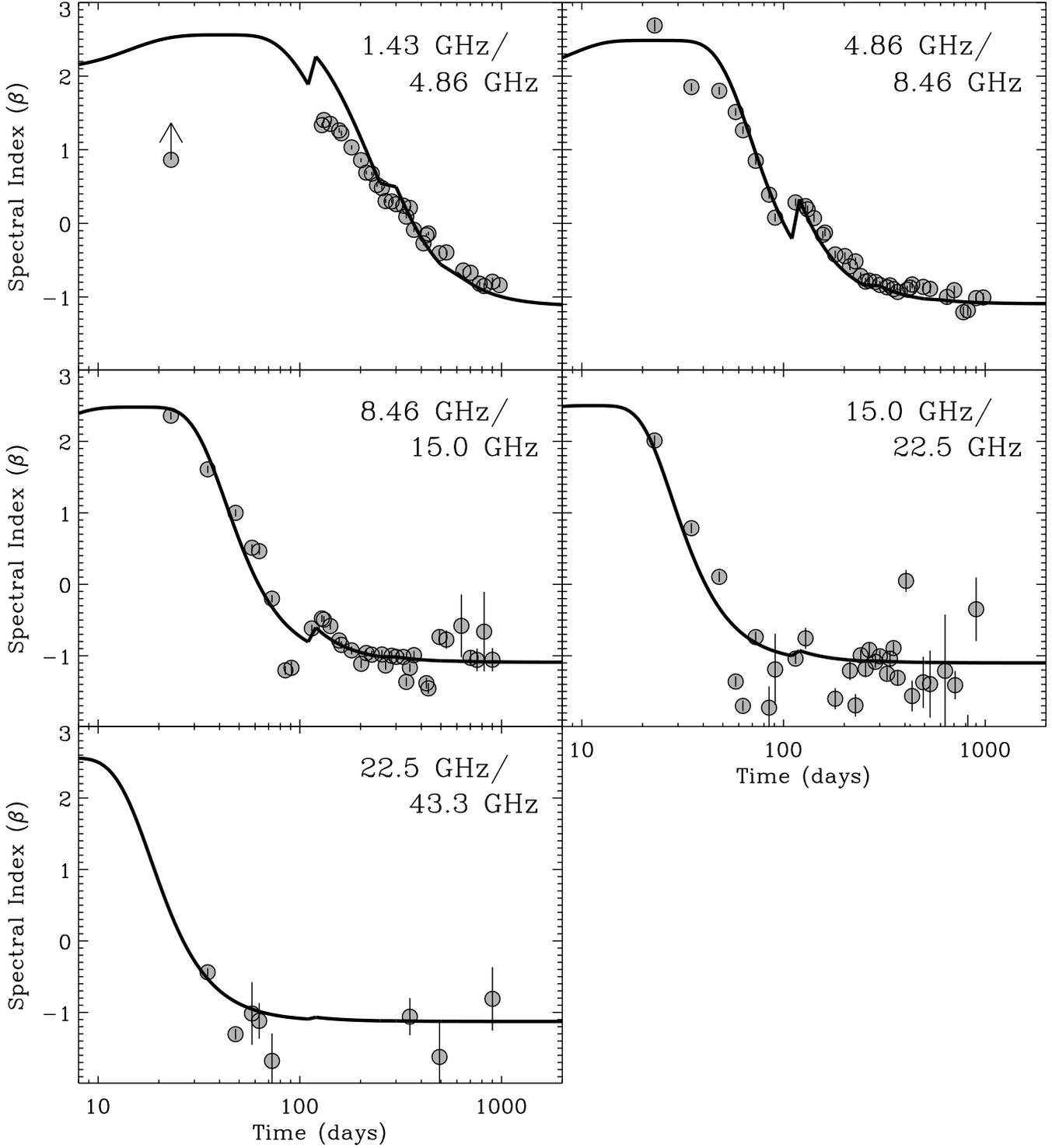}
\vspace{1cm}
\caption{Spectral indices for the radio emission of SN\,2003bg.  The
temporal evolution of 1.43/4.86 GHz, 4.86/8.46 GHz, 8.46/15.0 GHz,
15.0/22.5 GHz and 22.5/43.3 GHz spectral indices are shown along with
our SSA model fits (solid lines). The abrupt variations in the
spectral indices are overall consistent with our CSM density jump
model (\S\ref{sec:ssa}).
\label{fig:spectral_indices}}
\end{figure}

Shown in Figure~\ref{fig:spectral_indices} are the spectral indices
between each of the adjacent radio frequencies. A spectral turnover is
clearly observed for the lower frequencies at early time; at later
time the spectral slope converges to $\beta \approx -1.1$.  The
observed spectral shape is overall consistent with a non-thermal
synchrotron spectrum with a significant absorption component that
suppresses the low frequency emission.  In the case of radio
supernovae, a low frequency turn-over may be attributed to internal
synchrotron self-absorption (SSA; \citealt{c98}) or free-free
absorption (FFA; \citealt{wsp+86}) from the external medium.  The observed
spectral index of the optically-thick spectrum enables us to
distinguish between these two processes.  For radio supernovae
dominated by SSA, the radio spectrum is approximated by

\begin{equation}
F_{\nu}=1.582 F_{\nu_p} \left(\frac{\nu}{\nu_p}\right)^{5/2} \times \left( 1-{\rm exp}\left[-\left(\frac{\nu}{\nu_p}\right)^{-\frac{(5-2\beta_{\rm thin})}{2}}\right] \right),
\end{equation}

\noindent
where $\nu_p$ is the peak spectral frequency, $F_{\nu_p}$ is the flux
density at $\nu_p$, and $\beta_{\rm thin}$ is the optically-thin
spectral index \citep{c98}.  On the other hand, for FFA dominated
emission, the optically-thick radio spectrum steepens to

\begin{equation}
F_{\nu}=K_1 \left(\frac{\nu}{5~{\rm GHz}}\right)^{\beta_{\rm thin}} \times {\rm exp}\left[-K_2\left(\frac{\nu}{5~{\rm GHz}}\right)^{-2.1}\right]~~\rm mJy
\end{equation}

\noindent
where $K_1$ and $K_2$ are normalization constants \citep{wsp+86}. 

\begin{figure}
\vspace{-1cm}
\epsscale{0.8}
\plotone{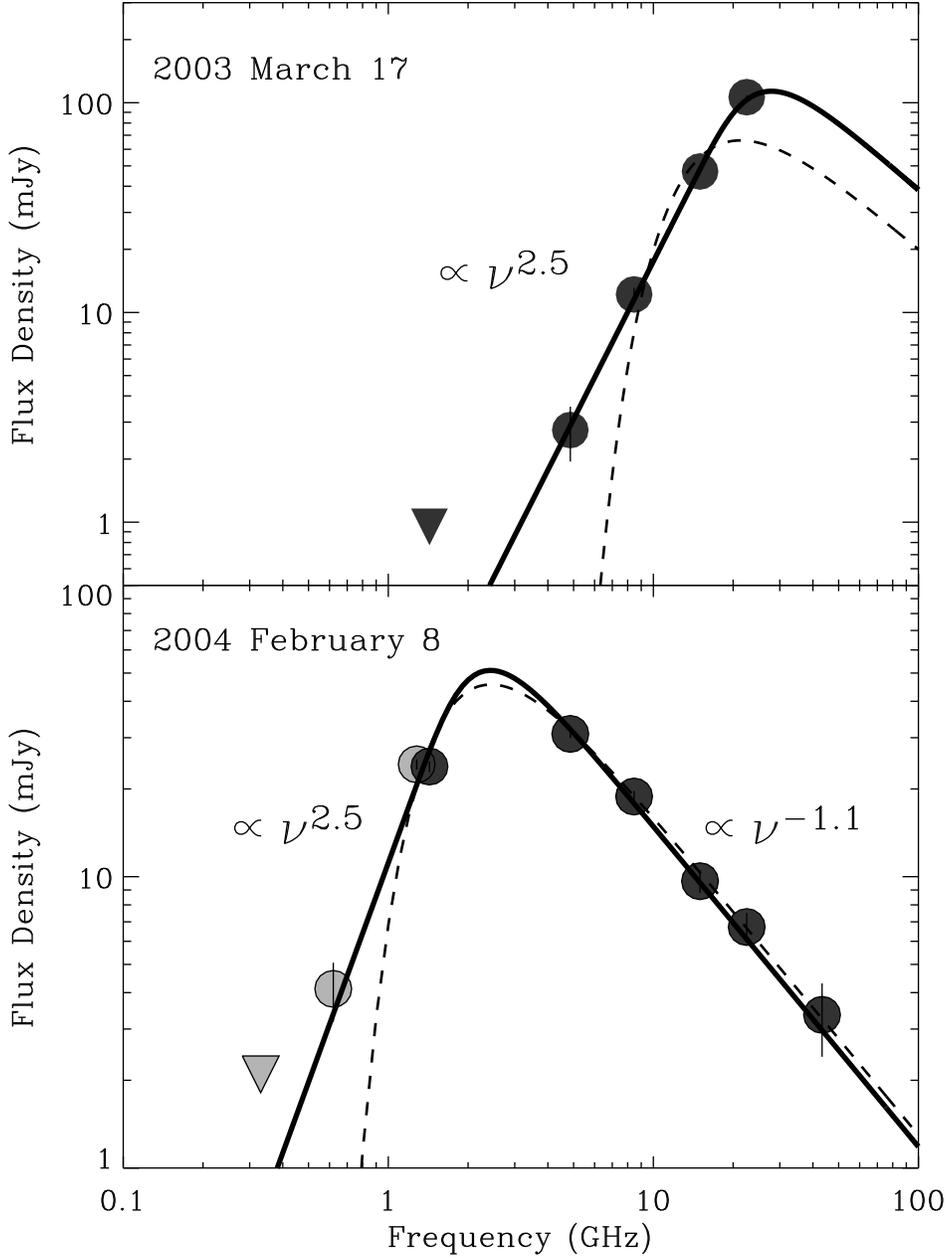}
\vspace{-1cm}
\caption{Radio spectra for SN\,2003bg on 2003 Mar 17 ($t\approx 23$
  days; top) and 2004 Feb 8 UT ($t\approx 351$ days; bottom).
  Detections are indicated by filled points and upper limits as
  inverted triangles. In the later spectrum, we supplement our Very
  Large Array data (dark grey points) with simultaneous observations
  from the Giant Meterwave Radio Telescope (light grey points;
  P. Chandra, 2005, PhD thesis,
  http://www.tifr.res.in/$\sim$poonam).  The low frequency GMRT
  observations at 0.33, 0.62 and 1.28 GHz confirm that the absorbed
  spectral index is not steeper than $\beta\approx 2.5$ \citep{cr05}.
  Both spectra are best-fit with a synchrotron self-absorbed spectrum
  (thick black lines) as described in \S\ref{sec:radio_spec}. For
  comparison, we show best-fit FFA models (dashed black lines) which
  are clearly inconsistent with the observed spectrum.  We conclude
  that internal SSA dominates the observed absorption.}
\label{fig:spectrum}
\end{figure}

We fit these functions to two well-sampled radio spectra collected on
2003 March 17 and 2004 February 8 ($t\approx 23$ and 351 days).  As shown in
Figure~\ref{fig:spectrum}, the optically-thick spectral component in
each epoch is best described with a spectral index of $\beta\approx
2.5$, and thus entirely consistent with the index predicted by the SSA
dominated model.  Specifically, we find the following best-fit SSA
parameters: $F_{\nu_p}\approx 110$ mJy (50 mJy) and $\nu_p\approx 25$
GHz (2.2 GHz) with $\chi^2_r\approx 1.5$ (4.5) for the early (late)
spectrum.  The FFA dominated models produce steeper optically-thick
spectra ($\beta > 2.5$) and thus provide a significantly worse fit to
the data.  We find the following best-fit FFA parameters: $K_1\approx
550$ (35) and $K_2\approx 11$ (0.12) with $\chi^2_r\approx 180$ (8.2)
for the early (late) spectra. Here we have adopted $\beta_{\rm
  thin}\approx -1.1$ for both models, consistent with the observed
optically-thin spectral indices from $t\sim 10$ to 1000 days
(Figure~\ref{fig:spectral_indices}).  Clearly the SSA model fit is
preferred over the FFA fit in both epochs.  We conclude that internal
SSA dominates the absorption on the timescale probed by the radio
data.

\subsubsection{Radio Polarization of SN\,2003bg}
\label{sec:pol}

Since synchrotron emission is inherently polarized, the magnetic field
geometry of radio SNe can be directly probed through polarization
measurements.  In order to measure the polarization for SN\,2003bg, we
carried out a full polarization calibration of our 8.46 GHz VLA
observations taken on 2003 October 7.4 UT ($t\approx 227$ days).  This run
was chosen for polarization calibration based on the high dynamic
range of the radio image.  To correct for the instrumental
polarization, we observed calibrator J0403+260 over a wide range in
parallactic angle and computed the leakage terms with AIPS task
PCAL.  We used 3C147 to calibrate the absolute polarization angle.

We do not find any significant polarization for SN\,2003bg.  In
particular, we place a limit on the linear polarization intensity
($I_{\rm pol}=\sqrt{Q^2+I^2}$, where $Q$ and $I$ are the Stokes
parameters) of $I_{\rm pol} < 0.071$ mJy, corresponding to a
fractional polarization limit of $\lesssim 0.8\%$ ($3\sigma$).  In
comparison with other radio SN linear polarization measurements this
is one of the deepest limits obtained to date: a factor of $\sim 4$
deeper than the limit for SN\,1993J ($\lesssim 3.3\%$, \citealt{bbr03})
and comparable to that for SN\,1979C ($\lesssim 1\%$,
\citealt{wsv+82}).

\subsection{X-ray Observations with \it Chandra\rm}
\label{sec:cxo}

SN\,2003bg was observed with the {\it Chandra} ACIS-S detector on 2003
March 24.5 UT ($t\sim 30$ days after the explosion) for 50.5 ksec.
\citet{pl03} reported the detection of an X-ray source at $\alpha\rm
(J2000)=04^{\rm h}10^{\rm m}59.42^{\rm s}, \delta\rm (J2000)=-31^{\rm
  o}24'50.3''$ ($\pm$0.5 arcsec in each coordinate), coincident with
the optical SN position.  The source was re-observed with ACIS-S on
2003 June 22.3 UT ($t\sim 120$ days) for 40.5 ksec to search for
variability between the two epochs.

We retrieved the SN\,2003bg data from the {\it Chandra} data
archive\footnote{http://cda.harvard.edu/chaser/} and reduced them
following the CIAO science
threads\footnote{http://cxc.harvard.edu/ciao/threads/}.  The source is
clearly detected in both epochs and we note that no diffuse emission
is detected from the host galaxy.  For spectral extraction of the SN
emission, we adopt a source aperture of 4.92 arcsec and a large
background region located $\sim 20$ arcsec from the source.  After
subtracting off the background emission, we measure count rates of
$0.0117\pm 0.0005$ and $0.0020\pm 0.0003$ cps (0.3-10 keV) in the
first and second epochs, respectively.

\citet{pl03} report that the first epoch data can be reasonably fit
with several different spectral models and propose a MEKAL hot plasma
model for the best fit.  Given that the small number of counts
prevents the model fits to be distinguished, we fit the extracted SN
spectra with only two basic models: absorbed power-law and thermal
bremsstrahlung.  Table~\ref{tab:cxo} lists the parameter values for
our resulting spectral fits, where we have adopted both fixed and
variable $N_{\rm H}$ values for comparison.  \citet{sfd+98} dust maps
give $E(B-V)=0.022$\, mag for the position of SN\,2003bg.  Using the
standard conversion of \citet{ps95}, we find $N_{\rm H}=1.3\times
10^{20}~\rm cm^{-2}$ for the Galactic column density along our the
line-of-sight.  The relatively few X-ray counts enable satisfactory
fits ($\chi^2_r\sim 1$) for each of the models.  

Given the significantly fewer counts in the second epoch, we fit the
spectrum with a fixed (Galactic) absorption.  Moreover, for the power-law
fit we adopt the best-fit spectral index from our first epoch.  The
resulting flux values appear in Table~\ref{tab:cxo} for both power-law and
thermal bremsstrahlung fits.  As in the case of the first epoch, 
both models provide an equally good representation of the data.

\begin{figure}
\plotone{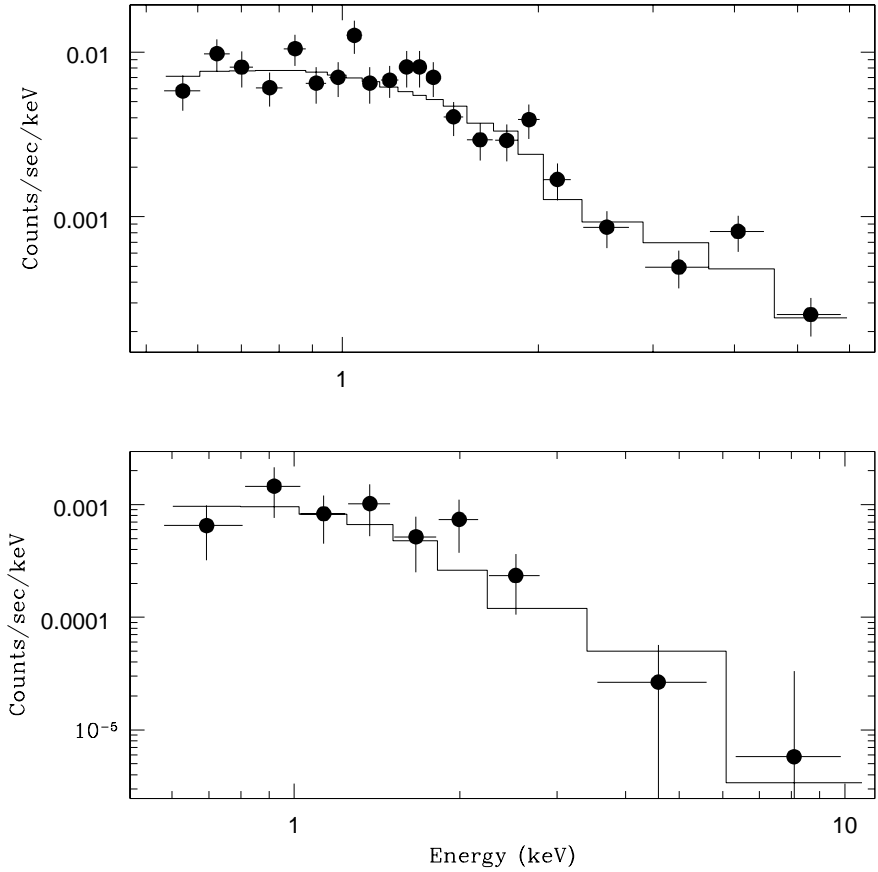}
\caption{{\it Chandra} ACIS-S spectra for SN\,2003bg at $t\approx 30$
  days (upper panel) and at $t\approx 120$ days (bottom panel).
  Overplotted in both panels is an absorbed power-law fit with
  fixed (Galatic) $N_{\rm H}$ as described in \S~\ref{sec:cxo} and
  Table~\ref{tab:cxo}.  In \S\ref{sec:xray_model} we argue that the
  X-ray emission is produced by inverse Compton scattering of the
  optical photons in the first epoch.
\label{fig:chandra}}
\end{figure}

In Figure~\ref{fig:chandra} we show the absorbed power-law model fits
for Epochs 1 and 2, assuming a fixed Galactic $N_H$.  Using the
associated spectral parameters in Table~\ref{tab:cxo}, we find
unabsorbed flux values of $F_X\approx (9.3\pm 0.4)\times 10^{-14}$ and
$(1.2\pm 0.4)\times 10^{-14}~\rm erg~cm^{-2}~s^{-1}$ (0.3--10 keV) for
these fits, respectively.

The observed temporal evolution is $F_{X}\propto t^{\alpha_X}$ with
$\alpha_X\approx -1.5\pm -0.3$ and the implied spectral indices
between the optically thin radio and X-ray band are nearly constant at
$\beta_{RX}\approx -0.94$ and $0.89$, for the first and second {\it Chandra}
epochs, respectively.  At a distance of 19.6 Mpc, the observed X-ray luminosity
values are thus $L_X\approx (4.3\pm 0.2)\times 10^{39}$ and
$(5.5\pm 1.9)\times 10^{38}~\rm erg~s^{-1}$, placing SN\,2003bg
among the most X-ray luminous SNe ever detected and a factor of $\sim 
10$ fainter than SN\,1998bw on a comparable timescale.

\section{Preliminary Constraints}
\label{sec:ep}

\citet{r94} showed that there is an upper limit to the brightness
temperature, $T_B$, for SSA dominated radio sources.  This limit
corresponds to the scenario where the fractions of post-shock energy
density in relativistic electrons ($\epsilon_e$) and magnetic fields
($\epsilon_B$) are in equipartition, and the shocked electrons are
accelerated into a power-law distribution, $N\propto \gamma^{-p}$,
above a minimum Lorentz factor, $\gamma_m$.  Under these assumptions,
the equipartition brightness temperature, $T_{\rm ep}\equiv
c^2 f_{\nu_p}/(2\pi k \theta_{\rm ep}^2 \nu_p^2)\approx 5\times
10^{10}$ K, defines the upper bound on $T_B$ \citep{r94,kfw+98}.
Here, $\theta_{\rm ep}$ is the equipartition size (radius) of the
radio source.  It is noted that $T_{\rm ep}$ depends only weakly on
the observed peak frequency and flux density.

Under the assumption that $T_B=T_{\rm ep}$, we can estimate the radius
and energy of the radio emitting material for SN\,2003bg.  Our first
VLA epoch in which the spectral turnover is observed is 2003 March 29
($t\sim 35$ days).  In this epoch we observed a peak flux density of
$f_{\nu_p} \approx 85$ mJy at ${\nu_p}\approx 22.5$ GHz and an
optically thin spectral index of $\beta=-(p-1)/2 \approx -1.1$
(Figure~\ref{fig:spectral_indices}).  As shown in
Figure~\ref{fig:spectrum}, on the timescale probed by our observations
the SN\,2003bg radio spectrum is well fit with a SSA model and we
therefore identify the spectral peak as the synchrotron
self-absorption frequency, $\nu_a$ (defined as the frequency where the
optical depth due to SSA is unity; $\tau_{\rm SSA}=1$). Using
Equations 2 and 3 of \citet{skb+05} (hereafter S05) the observed
source properties ($f_{\nu_p}$, $\nu_p$, $\beta$, and $d$) imply an
equipartition radius of $\theta_{\rm ep} \approx 40~\mu$as ($r_{\rm
ep} \approx 1.2\times 10^{16}$ cm) and a total energy for the radio
emitting material of $E_{\rm ep} \approx 1.7\times 10^{48}$ erg, where
we have assumed $\epsilon_e=\epsilon_B=0.1$.  The average velocity of
the radio shell is therefore roughly $\overline{v}_{\rm ep} \approx
0.13 c$ and the magnetic field is $B_{\rm ep} \approx 2.8
\epsilon_B^{1/2} E_{\rm ep}^{1/2}~r_{\rm ep}^{-3/2} \approx 0.9$ G.

A comparison with the ejecta velocities compiled for core-collapse SNe
\citep{c98} shows that SN\,2003bg resembles more closely a Type Ibc
(typical ejecta velocities, $\overline{v}\sim 0.1c$) rather than a
Type II event ($\overline{v}\sim 0.01c$).  Moreover, the inferred
ejecta velocity for SN\,2003bg is a factor of $\sim 10$ slower than
the mildly relativistic speed inferred for SN\,1998bw.  This analysis
emphasizes that broad optical absorption lines cannot be used as a
proxy for relativistic ejecta, consistent with the radio analysis for
broad-lined SNe 2002ap \citep{bkc02} and SN\,2003jd \citep{snk+05}.

Our preliminary constraint on the minimum energy for SN\,2003bg
places it among the most energetic radio SNe Ibc ever observed, second
only to SN\,1998bw and comparable to SN\,2003L (S05).  We note that
additional absorption processes (e.g. FFA) and departures from
equipartition (e.g. SN\,1993J; \citealt{fb98}) increase the energy
budget further.

Equipartition analysis may also be used to roughly constrain the
characteristic synchrotron frequency, $\nu_m$.  Equating particle
kinetic energy across the shock discontinuity, S05 show
that $\nu_m \approx 9.2\times 10^{3} \epsilon_e^2 (v/c)^4 B$ GHz which
gives $\nu_m \approx 0.02$ GHz for $\epsilon_e=0.1$ with the values of
$v_{\rm ep}$ and $B_{\rm ep}$ given above. At $t\sim 35$ days, $\nu_m$
is therefore already below our radio observing band and the synchrotron
break frequencies are ordered such that $\nu_m < \nu_a$.

Finally, we use these equipartition results at $t\approx 35$ days to
estimate the mass loss rate of the progenitor star.  From \citet{c98}, the
mass loss rate of the star, $\dot{M}$, can be derived from the
post-shock energy density in magnetic fields:

\begin{equation}
U_B=\frac{B^2}{8\pi}\approx \frac{\epsilon_B}{4\pi} \left(\frac{\dot{M}}{v_w}\right) r^{-2} v^{2}
\end{equation}

\noindent
where we have assumed a stellar wind-blown medium ($n\propto r^{-2}$).
We find $\dot{M}\approx 6.1\times 10^{-5}~\rm M_{\odot}~yr^{-1}$
(assuming a Wolf-Rayet wind velocity of $v_w=10^3~\rm km~s^{-1}$).

For comparison, we estimate the mass loss rate assuming FFA dominates
the absorption.  In this scenario, the shock radius is larger than the
equipartition estimate and so our equipartition parameters provide
only a lower limit on the mass loss rate of the progenitor star. Using
Equation 2.3 of \citep{flc96} we find:

\begin{equation}
\frac{\dot{M}_{-5}}{v_{w,3}} > 4.3\times 10^3 \tau_{\rm FFA}^{1/2} \frac{\nu_p}{\rm GHz} T_5^{3/4} \left(\frac{v}{c}\right)^{3/2} \left(\frac{t}{\rm 35~days}\right)^{3/2} 
\end{equation}

\noindent
where $\tau_{\rm FFA}$ is the optical depth to FFA processes and $T_5$
is the temperature normalized to $10^5$ K.  Here we have adopted the
notation $10^x Q_x=Q$ where $\dot{M}$ is given in units of $\rm
M_{\odot}~\rm yr^{-1}$ and $v_w$ is given in $\rm km~s^{-1}$.  Setting
$\tau_{\rm FFA}=1$ and $T_5=1$ we find $\dot{M}_{-5}/{v_{w,3}} >
4.5\times 10^3$ corresponding to a mass loss rate of $0.045~\rm
M_{\odot}~yr^{-1}$ ($v_w=10^3~\rm km~s^{-1}$) at $t\approx 35$ days.
This is three orders of magnitude larger than that derived from the
SSA interpretation.  Moreover, this would imply that the total energy
of the radio emitting material (a proxy for the kinetic energy of the
fastest ejecta, $v\gtrsim 0.13c$) exceeds $1.5\times 10^{51}$ erg.
Since hydrodynamic collapse distributes the ejecta kinetic energy as
$E_K\propto v^{-5}$ \citep{mm99}, this would imply that the optical
data (with $v\approx 0.06c$) have $E_K > 10^{53}$ erg, larger than any
other core-collapse SN to date.  This result, taken together with the
fact that a SSA model provides a significantly better fit to the radio
spectra, leads us to conclude that the observed spectral turn-over is
due to internal synchrotron self-absorption.

\section{Synchrotron Self-Absorption Model}
\label{sec:model}

In modeling the radio light-curves of SN\,2003bg, we adopt the
formalism of S05 where we presented a rigorous formulation of the
temporal and spectral evolution of synchrotron emission arising from
sub-relativistic supernova ejecta.  S05 show how the observed SN radio
emission spectrum at any single epoch is determined by three
parameters: $C_f$, $C_{\tau}$ and $\nu_{m}$.  Here, $C_f$ and
$C_{\tau}$ are normalization constants of the flux density and optical
depth with cgs units of ${\rm g~s^{1/2}}$ and ${\rm s^{-(2+p/2)}}$,
respectively.  The parameters $C_f$, $C_{\tau}$ and $\nu_{m}$ are in
turn determined by the values of four physical parameters: the
magnetic field, $B$, the shock radius, $r$, the minimum electron
Lorentz factor, $\gamma_{m}$, and the ratio
$\frak{F}\equiv\epsilon_e/\epsilon_B$.  With four physical parameters
($B$, $r$, $\gamma_{m}$, $\frak{F}$) and only three constraints
($C_f$, $C_{\tau}$, $\nu_{m,0}$), we must assume an additional
constraint to find a unique solution.  This constraint is obtained by
adopting a value for $\frak{F}$. By inverting the equations for $C_f$,
$C_{\tau}$ and $\nu_{m}$, S05 derive the following expressions for
$B$, $r$ and $\gamma_{m}$:

\begin{eqnarray}
B=9.0\times 10^{-8} ~(2+p)^{-6/17}(p-2)^{-4/17} \eta^{4/17} \frak{F}^{-1/17} \times \nonumber \\
\left(\frac{d}{\rm cm}\right)^{-4/17} \left(\frac{C_f}{\rm g~s^{1/2}}\right)^{-2/17} \left(\frac{C_{\tau}}{\rm s^{-(2+p/2)}}\right)^{4/17} \left(\frac{\nu_{m}}{\rm Hz}\right)^{-2(p-2)/17}~~\rm G
\label{eqn:B} \\
r=9.3\times 10^{12} ~(2+p)^{7/17} (p-2)^{-1/17} \eta^{1/17} \frak{F}^{-1/17} \times \nonumber \\
\left(\frac{d}{\rm cm}\right)^{16/17} \left(\frac{C_f}{\rm g~s^{1/2}}\right)^{8/17} \left(\frac{C_{\tau}}{\rm s^{-(2+p/2)}}\right)^{1/17} \left(\frac{\nu_{m}}{\rm Hz}\right)^{-(p-2)/34}~~\rm cm
\label{eqn:r}\\
\gamma_{m}=2.0 ~(2+p)^{3/17} (p-2)^{2/17} \eta^{-2/17} \frak{F}^{2/17} \times \nonumber \\
\left(\frac{d}{\rm cm}\right)^{2/17} \left(\frac{C_f}{\rm g~s^{1/2}}\right)^{1/17} \left(\frac{C_{\tau}}{\rm s^{-(2+p/2)}}\right)^{-2/17} \left(\frac{\nu_{m}}{\rm Hz}\right)^{(13+2p)/34}
\label{eqn:gamma_m}
\end{eqnarray}

\noindent
Here, $d$ is the distance to the supernova and $\eta$ characterizes
the the thickness of the radiating electron shell as $r/\eta$.  The
number density of the synchrotron emitting electrons then follows
directly as

\begin{equation}
n_e=\frac{p-2}{p-1} \frac{B^2}{8\pi} \frac{\frak{F}}{m_e c^2 \gamma_{m}}
\label{eqn:n_e}
\end{equation}

\noindent
where it is assumed that the contribution from electrons in a thermal
distribution is negligible.  The circumstellar density can be
expressed in terms of a progenitor mass loss rate

\begin{equation}
\dot{M}=\frac{8\pi}{\eta} n_e m_p r^2 v_w
\end{equation}
\label{eqn:m_dot}

\noindent
by adopting a value for the wind velocity.  The total energy of
the radio emitting material at a given epoch is then given by

\begin{equation}
E=\frac{4\pi}{\eta} r^3 \frac{\frak{F}}{\epsilon_{e}} \frac{B^2}{8\pi}
\label{eqn:energy}
\end{equation}

\noindent
where it should be noted that $E$ depends not only on $\frak{F}$ but
also on an assumed value of $\epsilon_e$.

\subsection{Hydrodynamical Evolution of the Ejecta}
\label{sec:hydro}

As discussed by \citet{c96}, several models have been proposed for
hydrodynamic evolution of the sub-relativistic supernova ejecta.
Based on separate assumptions, these analytic models constrain the
temporal behavior of the shock radius, magnetic field, electron
Lorentz factor, and energy partition fractions.  Here, we adopt the
standard SSA model (Model 1 of \citealt{c96}) for the hydrodynamical
evolution of the ejecta. This model assumes that the evolution is
self-similar across the shock discontinuity, and thus $r\propto
t^{\alpha_r}$ with $\alpha_r=(n-3)/(n-s)$ where $n$ characterizes the
density profile of the outer SN ejecta ($\rho \propto r^{-n}$) and $s$
characterizes the density profile of the radiating electrons within
the shocked circumstellar material ($n_e \propto r^{-s}$).  In
addition, the standard SSA model assumes that the magnetic energy
density ($U_B \propto B^2$) and the relativistic electron energy
density ($U_e \propto n_e \gamma_m$) scale as the total post-shock
energy density ($U \propto n_e v^2$).  In this scenario, the magnetic
field is amplified by turbulence near the shock discontinuity,
implying fixed energy fractions, $\epsilon_e$ and $\epsilon_B$, and
thus a constant value of $\frak{F}$ throughout the evolution of the
ejecta.

The evolution of the magnetic field is determined by the CSM density
profile and the expansion of the shock radius: $B\propto t^{\alpha_B}$
with $\alpha_B=\alpha_r(2-s)/2-1$.  We note that for a wind stratified
medium, $s=2$ and $\alpha_B=-1$.  The minimum Lorentz factor evolves
as $\gamma_m\propto t^{\alpha_{\gamma}}$ with
$\alpha_{\gamma}=2(\alpha_r-1)$.  From the scalings of
Equation~\ref{eqn:energy} it follows that $E\propto t^{3\alpha_r -2}$,
and therefore the total post-shock energy increases with time until
the swept up circumstellar mass is comparable to rest mass of the SN
ejecta.

\section{SSA Model Fit for SN\,2003bg}
\label{sec:ssa}

As shown in Figure~\ref{fig:oneplot} and briefly discussed in
\S\ref{sec:obs}, the radio light-curves for SN\,2003bg are
characterized by achromatic short timescale variability, inconsistent
with the strict power-law evolution prescribed by the standard model.
In an effort to model the overall evolution of the radio ejecta we
therefore divide the multi-frequency light-curves into four
subsections, each defined by the observed time of abrupt variation.
We then apply the SSA model described in \S\ref{sec:model} to each
subsection.

\subsection{The Synchrotron Self-absorption Peak}
\label{sec:first}
\begin{figure}
\plotone{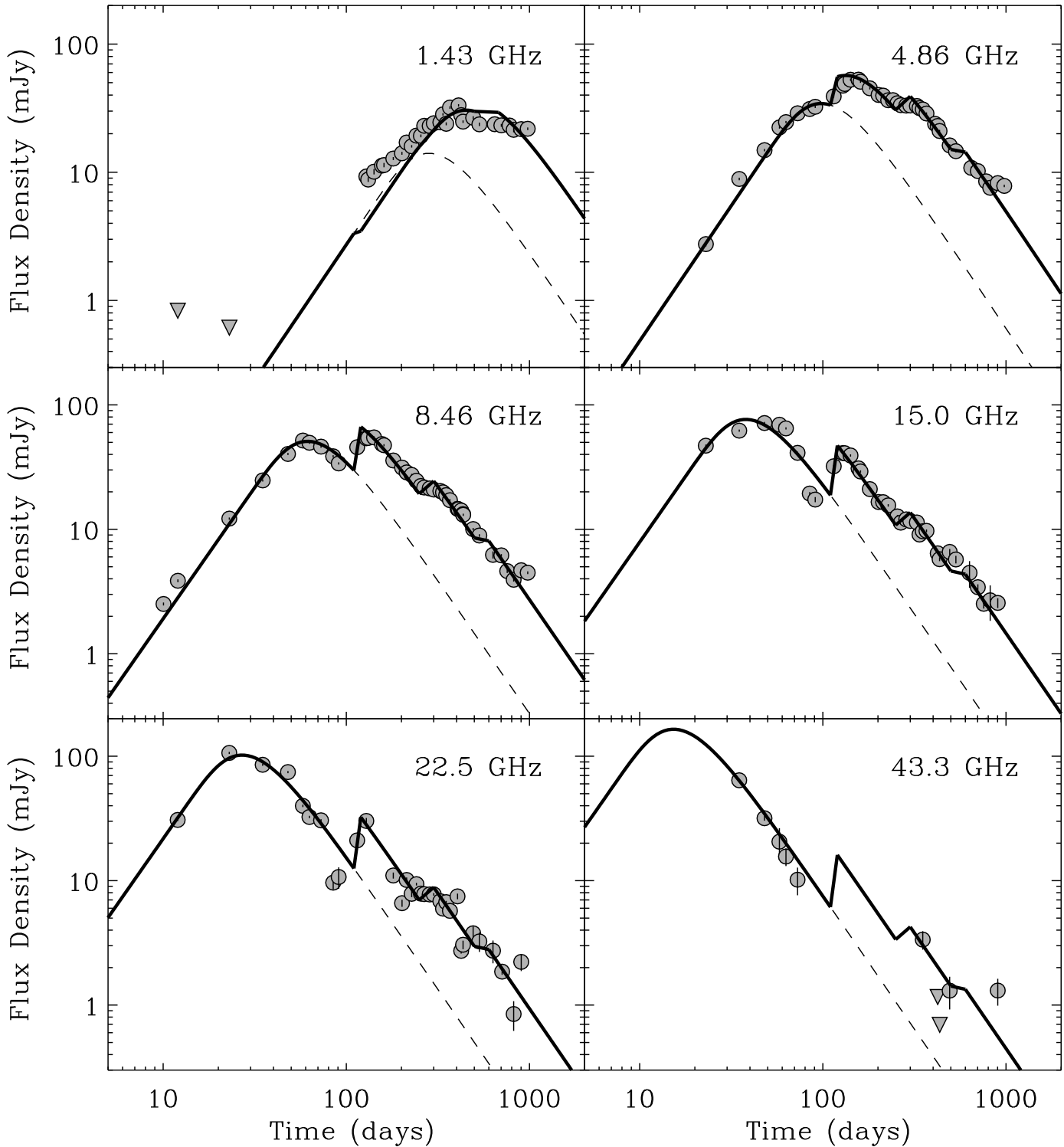}
\vspace{1cm}
\caption{Radio light-curves of SN\,2003bg were taken
  with the VLA at frequencies 1.43, 4.86, 8.46, 15.0, 22.5 and 43.3 GHz
  between 2003 March and 2005 October UT.  The synchrotron
  self-absorption model fits described in \S\ref{sec:ssa} are
  overplotted (solid line).  An extrapolation of our SSA model
  with no density variations is shown for comparison (dashed line).
\label{fig:lt_curves}}
\end{figure}

Figure~\ref{fig:lt_curves} shows that the first peak observed for
each of the radio frequencies is chromatic, and therefore produced by
a cascading spectral break.  We define the time range for
this chromatic evolution phase to be $t < 110$ days, extending until just
before the first achromatic variation is observed.  As discussed in 
\S\ref{sec:obs} the radio emission is well described by a SSA dominated
spectrum and we therefore attribute this chromatic subsection to 
the passage of $\nu_a$ through the observed frequencies.
The evolution of the spectral indices
(Figure~\ref{fig:spectral_indices}) indicate that the self-absorption
frequency cascades as $\nu_a\propto t^{-1.1}$, consistent with typical
values observed for other SSA dominated radio supernovae \citep{c98}.

Using our multi-frequency radio data collected during the time range
of the first peak and adopting $\frak{F}=1$, we fit for the constants
$C_f$ and $C_{\tau}$ as well as the temporal indices $\alpha_r$ and
$\alpha_B$.  As discussed in Section~\ref{sec:ep}, $\nu_m$ is
estimated to be below the radio band during our observations and thus
is unconstrained by the data.  We therefore estimate $\nu_{m,0}\approx
0.1$ GHz at our chosen reference time of $t_0=10$ days\footnote{Here
(and throughout) we use subscript ``0'' to denote the values of
parameters at the reference time.}, consistent with the preliminary
value calculated in \S\ref{sec:ep}.  We find a best-fit solution
($\chi^2_r\approx 46$; dominated by interstellar scintillation) for
parameters: $p\approx 3.2$, $C_f\approx 3.1\times 10^{-51}~\rm g~s^{1/2}$,
$C_{\tau}\approx 6.2\times 10^{38}~\rm s^{-3.6}$, $\alpha_r\approx 0.8$ and
$\alpha_B\approx -1$.  These values imply $s\approx 2$, $n\approx 7$
and $\alpha_{\gamma}\approx -0.4$.  As explained within the Appendix of S05,
we parameterize the sharpness of the $\nu_a$ spectral break with
$\zeta=[0,1]$ and find $\zeta\approx 0.6$ for our final solution.
This SSA model provides a reasonable fit to the chromatic subsection
of the light-curves as shown in Figures~\ref{fig:oneplot} and
\ref{fig:lt_curves}.

With this SSA model fit, the physical parameters of the ejecta and CSM
are uniquely determined by Equations~\ref{eqn:B} - \ref{eqn:energy}.
\citet{c82a} show that for $s\approx 2$ and $n\approx 7$, $\eta\approx
4$. Adopting this value, we find that the expansion of the ejecta is
described by $r\approx 6.2\times 10^{15} (t/t_0)^{0.8}$ cm, and the
average velocity is $\overline{v}\approx 0.24 (t/t_0)^{-0.2} c$.  The
radio ejecta therefore expand with a modest sub-relativistic velocity
comparable to that observed for SN\,2002ap \citep{bkc02} and
significantly slower than SN\,1998bw (bulk Lorentz factor, $\Gamma\sim 
2$ on a similar timescale \citep{kfw+98,lc99}.  Adopting
$\epsilon_e=\epsilon_B=0.1$ we find that the total energy of the radio
emitting material is given by $E\approx 7.3\times 10^{48}
(t/t_0)^{0.4}$ erg, comparable to that of SN\,1998bw.  The magnetic
field evolves radially as $B\approx 4.9 (r/r_0)^{-1.25}$ G, slightly
steeper than that observed for Type Ic SNe 2003L (S05), 2002ap
\citep{bkc02} and the Type IIb SN1993J \citep{fb98}. The electron
number density and associated mass loss rate are thus given by
$n_e\approx 2.2\times 10^{5} (r/r_0)^{-2}~\rm cm^{-3}$ and
$\dot{M}\approx 1.4\times 10^{-4}~\rm M_{\odot}~yr^{-1}$ where we have
assumed a Wolf-Rayet wind velocity of $v_w=1000~\rm km~s^{-1}$ and
adopted a nucleon-to-electron density ratio of two (appropriate for WR
winds).  This mass loss rate is consistent with the values observed
for Galactic Wolf-Rayet stars \citep{cgv04}.

\subsection{Subsequent Light-curve Variations}
\label{sec:second}

Shortly after the observed SSA peak cascades through 8.46 GHz, an
abrupt rise was observed at the optically thin frequencies. By $t\sim 
120$ days the radio emission reached an {\it achromatic} second peak.
Subsequent achromatic light-curve variations were later observed at
$t\sim 300$ and 600 days and are most pronounced at frequencies above
$\nu_a$.  As shown in Figure~\ref{fig:spectral_indices} these
variations were associated with abrupt changes in the spectral
indices, most notably for those indices which straddle $\nu_a$.

Possible causes for these abrupt light-curve variations include energy
injection from a central engine, interaction of the reverse shock with
density structures in the ejecta, and interaction of the forward shock
with CSM density enhancements. \citet{kfw+98} invoke energy injection
to explain the achromatic second peak in the radio light-curves of
SN\,1998bw at $t\sim 40$ days.  This interpretation was supported by
the inferred mildly relativistic ejecta speeds and associated prompt
gamma-ray emission, suggestive that SN\,1998bw was powered by a
GRB-like central engine.  In the case of SN\,2003bg, however, the
inferred ejecta speed is merely sub-relativistic and therefore there
is no evidence for an engine-driven explosion.  We conclude that the
variations observed for SN\,2003bg cannot be attributed to long-lived
engine activity.

\citet{mdb01} show that the late-time flattening observed for the
SN\,1993J radio light-curves can be produced through the interaction
of the reverse shock with density inhomogeneities within the ejecta.
While the 1-D hydrodynamical simulations provide a reasonable fit to
the smooth evolution of SN\,1993J at $t\gtrsim 1000$ days, it is not
clear that they could produce the abrupt rise at $t\sim 100$ days 
observed for SN\,2003bg.  Moreover, any abrupt features in the ejecta would
likely be smoothed by instabilities in a 3-D simulation.

In the third scenario, the interaction between the forward shock and
strong density variations in the circumstellar material produces
abrupt changes in the temporal evolution of the radio light-curves.
For a wind-stratified medium with constant progenitor mass loss rate,
the circumstellar density is predicted to follow a smooth $r^{-2}$
profile.  Abrupt deviations from a wind density profile have
been inferred for several radio supernovae (e.g. SN\,1979C;
\citealt{wvd+91}) and are argued to be the result of a variable
mass loss history and/or the effects of a binary companion on the
structure of the CSM.  These density induced variations are more
pronounced at optically thin frequencies than for optically thick.
Here we show that the light-curve variations observed for SN\,2003bg
are similarly due to CSM density enhancements.

We model the SN\,2003bg radio light-curves with a simple density
enhancement model under the following assumptions: (a) the density
variations can be approximated by modest abrupt jumps in the CSM such
that the density profile returns to $n_e\propto r^{-2}$ after each
enhancement, (b) the hydrodynamical evolution of the ejecta returns to
the self-similar solution after each jump, thereby maintaining the
same values of $\alpha_r$, $\alpha_B$ and $\alpha_{\gamma}$ both
before and after each density enhancement, (c) the microphysics of the
shock are not affected by the density jumps and therefore $\epsilon_e$
and $\epsilon_B$ remain constant throughout the evolution of the
ejecta, (d) the bulk motion of the ejecta is not significantly
affected by the density jumps since the slower, more massive material
behind the shock continues to plow forward, and (e) the minimum
Lorentz factor of the radiating electrons is not affected by the CSM
enhancements, although the number of electrons participating in the
radio emission increases.

Maintaining a self-similar evolution throughout the density variations
requires that the magnetic field is enhanced by each of the CSM jumps.
Moreover, since $\epsilon_e$ and $\epsilon_B$ are assumed to be fixed
constants, the additional thermal energy produced by these
enhancements causes an increase in the total energy of the radio
emitting region.  Therefore, in fitting the subsequent peaks in the
SN\,2003bg radio light-curves, we assume that the evolution of $r$ and
$\gamma_m$ remain effectively unchanged while $B$ and $E$ vary
according to the strength of the CSM density enhancements.

We recognize that this simple density enhancement model is not fully
consistent with the hydrodynamic evolution of the ejecta.  In fact, a
small change in the radial expansion is expected as the shock wave
reaches the higher density region. For the $n\approx 7$ self-similar
solution adopted here and the modest density enhancements we infer, we
estimate this to be minor effect ($\sim 10\%$).  Therefore, the
variations in ejecta parameters ($B$ and $E$) are primarily the result
of the CSM density enhancements rather than the minor variations in
radial expansion.  However, the series of density jumps that we infer
cause the evolution of the radio ejecta to gradually deviate from the
initial radial evolution, making our model less accurate at very late
time (after several years).  In addition, we note that the outflow
does not immediately settle on the new self-similar solution following
each density variation, but instead settles on a timescale comparable
for the radius to double (i.e. the ratio of the settling time to the
age of the blastwave is of order unity).  Thus the sharp features
appearing in our light-curve fits are merely an artifact of our model
assumptions.

As discussed in \S\ref{sec:model}, $C_f$, $C_{\tau}$ and $\nu_m$ are
functions of just three physical parameters: $r$, $\gamma_m$ and $B$.
Of these, only $B$ responds significantly to the CSM density
variations.  From the Appendix of S05 we have $C_f\propto B^{-1/2}$,
$C_{\tau}\propto B^{3+p/2}$ and $\nu_m\propto B$.  By fitting for $B$
at each density variation, the new values of $C_f$, $C_{\tau}$ and $\nu_m$
are thus directly determined.

The observed timescales and estimated circumstellar radii for each
density enhancement are given in Table~\ref{tab:peaks}.  We note that
we do not fit the short rise time of each phase during which the
ejecta are settling to the new self-similar solution. As shown in
Figure~\ref{fig:lt_curves}, we find reasonable fits for each of the
light-curve phases by invoking modest (factor of $\lesssim 2$) density
enhancements.  In Table~\ref{tab:peaks} we list the values of
$C_f$, $C_{\tau}$ and $\nu_m$ and the associated enhancement factors
for the density, magnetic field and ejecta energy for each phase.  It
is interesting to note that the strength of the enhancements decay
with time.

\section{Radio Polarization}
\label{sec:pol_disc}

As discussed in in \S\ref{sec:obs}, our VLA observations constrain the
linear polarization of the radio emission to less than $\sim 0.8\%$.
Although the synchrotron emission from the radio ejecta is inherently
polarized, the observed polarization level can be significantly
suppressed for two reasons.  First, since the SN\,2003bg radiosphere
is not resolved in the VLA data our polarization limit applies to the
{\em integrated} polarization of the ejecta emission.  As shown for
supernova remnant Cas A, the integrated polarization can be
significantly suppressed due to the approximate circular symmetry of
the radio ejecta \citep{r70}.  Second, the radio emission can be
depolarized due to internal Faraday dispersion within the ejecta
\citep{b66,c82b}.  For the density and magnetic field values we derive
from our SSA model, we predict the integrated linear polarization of the
SN\,2003bg radio emission to be negligible, consistent with the
observations.

\section{Modeling the X-ray emission}
\label{sec:xray_model}

X-ray emission in supernovae can be produced by three processes
\citep{flc96}: [1] non-thermal synchrotron emission from radiating
electrons, [2] thermal (free-free) bremsstrahlung emission from
material in the circumstellar shock and/or the ejecta reverse shock,
and [3] inverse Compton scattering of photospheric photons by
relativistic electrons.  We examine each of these scenarios in the
context of the bright X-ray emission observed for SN\,2003bg.

\subsection{Synchrotron Emission}
As shown in Table~\ref{tab:cxo}, the spectral index between the
optically thin radio and X-ray emission is $\beta\approx -0.9$ for
both {\it Chandra} epochs.  Assuming a constant injection spectral
index from the radio to X-ray bands we extrapolate the observed radio
spectral index of $\beta\approx 1.1$ to the X-ray
and find that the synchrotron emission under-predicts the observed
X-ray flux by a factor of $\sim 10$ in both epochs.  This discrepancy
grows significantly larger when the synchrotron cooling break,
$\nu_c=18\pi m_e c e/(t^2\sigma_T^2 B^3)$, is included, beyond
which the spectrum steepens by $\Delta\beta=-0.5$.  Using the magnetic
field derived in Section~\ref{sec:ssa}, the cooling frequency at
$t\approx 30$ (120) days is $\nu_c \approx 6.1 \times 10^{10}$
($2.3\times 10^{11}$) Hz.  Including the cooling break and
extrapolating from $\nu_c$ to $\nu_{\rm X-ray}$ as $f_{\nu} \propto
\nu^{-1.6}$, the synchrotron flux falls five orders of magnitude below
the observed X-ray emission.  We therefore conclude that the X-ray
emission is not dominated by synchrotron emission.

\subsection{Thermal Bremsstrahlung Emission}
In a scenario dominated by thermal bremsstrahlung processes, the X-ray
emission is produced by the forward shock plowing into circumstellar
material and/or the reverse shock heating of the ejecta.  In this
case, the strength of the X-ray emission depends strongly on the
density of the CSM and unshocked ejecta.  The shocked material then
cools by free-free emission processes.  For a wind stratified density
profile, the summed free-free luminosity from the forward and reverse
shock is generalized \citep{cf01,scb+03} roughly by

\begin{equation}
L_{X,\rm FS+RS}\approx 8.6\times 10^{34} C_{L,\rm FS+RS} \dot{M}_{-5}^2 v_{w,3}^{-2} t_1^{-1}~~\rm erg~s^{-1}
\end{equation}

\noindent
where $C_{L,\rm FS+RS}$ is a constant such that $C_{L,\rm
  FS+RS}=1+(n-3)(n-4)^2/(4n-8)$.  From our SSA radio model we find
  $\alpha_r\approx 0.8$ and $\dot{M}_{-5}\approx 14$ ($v_{w,3}=1$),
  thus $C_{L,\rm FS+RS}\approx 2.8$ and $L_{X,\rm FS+RS}\approx
  1.6\times 10^{37}~\rm erg~s^{-1}$.  The predicted X-ray luminosity
  is therefore a factor of $\sim 10^2$ fainter than the first epoch
  observation.

On the other hand, if we assume the X-ray emission must be dominated
by free-free processes, then this implies a significantly higher mass
loss rate of $\dot{M}_{-5}\approx 230$ ($v_{w,3}=1$).  However, we
reiterate that a high density model dominated by external absorption
processes is inconsistent with the observed radio spectrum
(Figure~\ref{fig:spectrum}).  A possible compromise could be that the
X-ray emission is produced by dense CSM clumps distributed with a very
low filling factor ($f<<1$) such that the radio emission is optically
thin.  This scenario was invoked for the radio luminous Type IIn
SN\,1986J \citep{wps90}. However, the dense CSM clumps inferred for
Type IIn's also give rise to bright nebular H$\alpha$ emission lines
which are not present in the nebular spectra of SN\,2003bg.  Moreover,
it would be difficult to sustain the ejecta velocities inferred from
our radio modeling in such a a high density environment (Chevalier \&
Fransson, {\it in prep}).  We conclude that the radio and optical
emission do not support a scenario in which the observed X-ray flux is
dominated by thermal bremsstrahlung emission.

\subsection{Inverse Compton Scattering}
X-ray emission from supernovae can also be produced by inverse Compton
scattering (ICS) of the relativistic electrons by the optical photons.
This process was observed to be important in explaining the observed
X-ray emission for SN\,2002ap \citep{bf04} and SN\,2003L (S05) on
timescales comparable to the optical maximum.  In this scenario the
ratio of flux measured in the radio and X-ray bands, respectively is
given by

\begin{equation}
\frac{F_{\rm radio}}{F_X}\approx \frac{U_B}{U_{\rm ph}}
\label{eqn:fxfr}
\end{equation}

\noindent
where $U_B$ and $U_{\rm ph}$ are the energy densities in magnetic
fields and up-scattered photons, respectively.  We note that
Equation~\ref{eqn:fxfr} is relatively insensitive to the spectral energy
index of the electrons.  \citet{bf04} show that

\begin{equation}
U_{\rm ph}\approx 0.4~L_{\rm bol,42} \left(\frac{t_d v}{c}\right)^{-2}~\rm erg~cm^{-3}
\label{eqn:uph}
\end{equation}

\noindent
where $t_d$ is the time in days, $L_{\rm bol, 42}$ is the optical
luminosity at $t=t_d$ normalized to $10^{42}~\rm erg~s^{-1}$ and $v$
is the shock velocity.  

At the time at the first X-ray observation, $t_d\approx 30$, the
optically thin radio flux was $F_{\rm radio}\approx 2.2\times
10^{-14}~\rm erg~cm^{-2}~s^{-1}$ and the unabsorbed X-ray flux was
$F_X\approx 9.3\times 10^{-14}~\rm erg~cm^{-2}~s^{-1}$; therefore
$F_{\rm radio}/F_X\approx 0.24$.  From our SSA radio modeling
(\S\ref{sec:ssa}) we estimate $B\approx 1.6$ G and $v\approx 0.19c$ at
$t\approx 30$ days.  Based on the optical peak magnitude (Hamuy et
al., private communication) we find $L_{\rm bol,42}\approx 2.3$ at
$t_d\approx 30$, coincident with the time of optical maximum.
Therefore, $U_B=B^2/8\pi \approx 0.10~\rm erg~cm^{-3}$, and $U_{\rm
ph}\approx 0.028~\rm erg~cm^{-3}$.  Thus $U_B/U_{\rm ph}\approx 0.28$,
comparable to the observed ratio for $F_{\rm radio}/F_X$.  We conclude
that inverse Compton emission can produce the bright X-ray luminosity
in the first epoch.

The second epoch, however, cannot be explained through ICS.  From our
radio model at $t_d\approx 120$, $B\approx 0.41$ G, $v\approx 0.15c$ and
$F_{\rm radio}\approx 1.5\times 10^{-15}~\rm erg~cm^{-2}~s^{-1}$.  The
optical light-curve implies $L_{\rm bol,42}\approx 0.23$ at this time.
Therefore we estimate $U_B/U_{\rm ph}\approx 24$ and a predicted IC
X-ray flux of $6.3\times 10^{-17}~\rm erg~cm^{-2}~s^{-1}$ which is a 
factor of $\sim 200$ fainter than that observed.  We conclude that
the X-ray flux observed at 120 days is dominated by a separate
process, possibly due to a cosmic ray dominated SN shock which
produces a late-time flattening of X-ray light-curves in SNe Ibc
(Chevalier $\&$ Fransson, {\it in prep.})
 
\section{Radio Supernovae and CSM Density Variations}
\label{sec:bumps}

We have shown that the abrupt radio light-curve variations observed
for SN\,2003bg can be attributed to density enhancements in the
circumstellar medium.  In comparison with well-sampled radio
supernovae compiled from previous
studies\footnote{http://rsd-www.nrl.navy.mil/7213/weiler/sne-home.html}
including our ongoing survey of Type Ibc events (Soderberg {\it et
  al.}, in prep), we find that $\sim 50\%$ of radio supernovae
similarly show evidence for abrupt achromatic variations. At the same
time, there are several examples of well-studied radio supernovae with
smooth light-curve evolutions including Type IIb SN\,1993J
\citep{vws+94,bbr03} and Type Ic SN\,2003L (S05) among others.  In
Table~\ref{tab:bumps}, we compile the light-curve properties for radio
supernovae with abrupt variability.  We emphasize that this
compilation contains core-collapse SNe from nearly all spectroscopic
classifications, including hydrogen-rich Type IIL, IIn and IIb in
addition to hydrogen-poor Type Ibc.

Table~\ref{tab:bumps} shows that the short timescale variability
observed for radio supernovae can be qualitatively divided into three
types.  The most common effect is an abrupt steepening of the radio
light-curves as observed for SN\,1988Z \citep{wpv+02}.  On the other
hand, an abrupt rise in the optically thin flux has been observed for
a couple of events (e.g. SN\,1987A, \citealt{mgw+02}).
More intriguingly, episodic (and perhaps periodic) light-curve bumps
are observed for SNe 1979C \citep{wvp+92,mvw+98,mwv+00}, 2001ig
\citep{rss+04}, 1998bw \citep{kfw+98,wpm01} and 2003bg, all of which
are characterized by modest flux deviations (factor of $\lesssim 2$).

These three types of radio variability are generally attributed to
dynamical interaction of the ejecta with irregularities in the
circumstellar medium.  Abrupt density variations in the media
surrounding core-collapse supernovae are known to be produced in a
variety of ways, including: [1] fluctuations in the progenitor mass
loss rate and/or wind velocity.  This scenario is supported through
observations of local supernova remnants and massive stars which
collectively suggest a complex picture for mass loss during the final
stage of stellar evolution. [2] Termination shocks between stellar
winds ejected during the evolution of the progenitor star may produce
abrupt density enhancements at radii of $R\gtrsim 0.1$ pc
\citep{glm96,gml96,clf04,rgs+05}.  Case 2 is observationally motivated
by the circumstellar nebulae associated with local Wolf-Rayet stars,
which are produced by interacting stellar winds
\citep{chu02}.  [3] a clumpy stellar wind perhaps resulting from
fragmentation of circumstellar shells by Rayleigh-Taylor instabilities
\citep{glm96}.  This scenario is motivated by narrow H$\alpha$ line
profiles observed for Type IIn supernovae. In the case of SN\,1988Z,
\citet{cd94} propose a shocked dense wind component at $R\sim 5\times
10^{16}$ cm to explain the low velocity line emission.  [4] the
effects of a binary companion on the circumstellar wind profile.  This
case is observationally motivated by the binary companion for Type IIb
SN\,1993J \citep{msk+04} and the dense circumstellar ring around
SN\,1987A at $R\sim 8\times 10^{17}$ cm \citep{bkh+95} commonly
attributed to the effects of a companion star \citep{pod02}.  In this
scenario, the binary system parameters determine the circumstellar
radii at which the progenitor wind is most affected.  [5] a close
binary progenitor system embedded in a common envelope \citep{pjh92}.
Given that the rate of SN progenitors in common envelope
binary systems is predicted to be low ($\sim 1\%$ of SNe Ibc;
\citealt{chu97}) this scenario is expected to be rare.  As an example,
a common envelope system has been argued to explain the large inferred
density enhancement at $R\sim 7\times 10^{16}$ cm responsible for strong
radio \citep{svs+04}, X-ray \citep{pl04} and optical (H$\alpha$;
\citealt{sgk04}) emission in the unusual Type Ic SN\,2001em at $t\sim 
2.5$ years \citep{cc05}.

To summarize, the strength and location of circumstellar density
variations depends on the method by which the density profile was
disturbed.  We note that Cases 3 and 5 are generally associated with
hydrogen-rich events whereas Cases 1, 2 and 4 are applicable to all
classes of core-collapse supernovae.

Finally, in light of this compilation it is interesting to note that
an abrupt rise in the optically thin radio emission of Type Ibc
supernovae can also be attributed to the presence of an off-axis
gamma-ray burst jet \citep{pac01,wax04}.  As the jet sweeps up
circumstellar material and decelerates, it eventually undergoes a
dynamical transition to sub-relativistic expansion on a timescale of
$t\sim 1$ yr \citep{fwk00}. Once sub-relativistic, the jet spreads
sideways and the ejecta rapidly approach spherical symmetry.  The
observational signature of an off-axis GRB is therefore an abrupt
optically thin rise as the jet sweeps through our line-of-sight.  Due
to the location of the synchrotron spectral peak, this signature is is
most easily detected in the radio band on a timescale of $t\sim 1$ to
10 years \citep{snk+05}.  Yet as we have shown here, many radio
supernovae show evidence for density induced variability on this same
timescale.  We therefore emphasize that abrupt flux variations in
radio supernovae may not be used as a reliable proxy for an off-axis
GRB jet and/or energy injection from a central engine.

\subsection{A Comparison of SN\,2003bg and SN\,2001ig}
\label{sec:comp}

\begin{figure}
\vspace{-1cm}
\epsscale{0.8}
\plotone{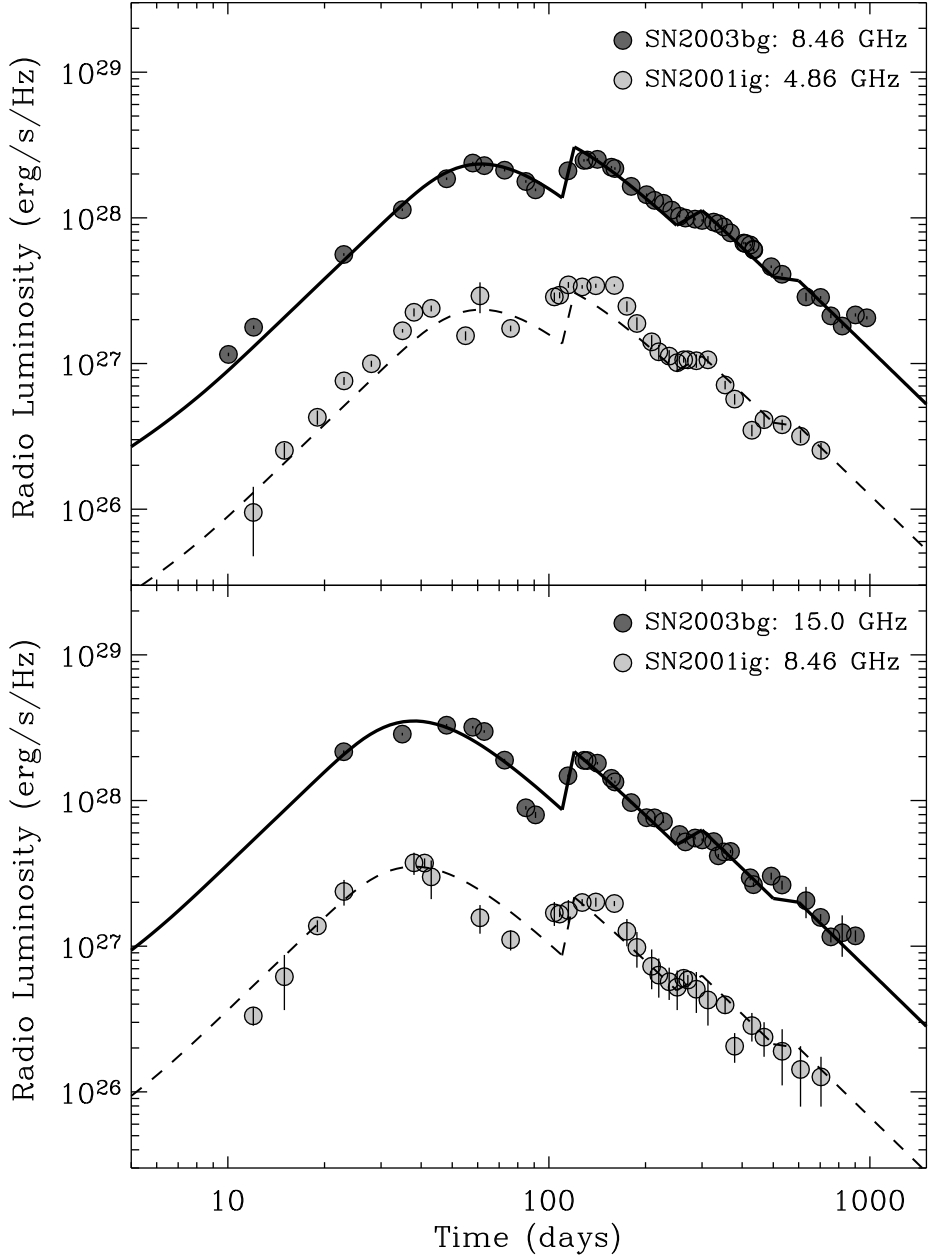}
\vspace{-1cm}
\caption{Radio light-curves for SN\,2003bg (dark grey circles) and the
  Type IIb SN\,2001ig (light grey circles; \citealt{rss+04}) are
  compared.  While SN\,2003bg is overall a factor of $\sim 10$ more
  luminous than SN\,2001ig, the shape of the radio light-curves are
  strikingly similar.  Specifically, the SNe show evidence for
  light-curve variations on similar timescales, perhaps indicative of
  similar CSM density structures and progenitor system evolution.  Our
  radio modeling fits (thick black lines) for SN\,2003bg are shown at
  8.46 (top panel) and 15.0 GHz (bottom panel).  By dimming the SN\,2003bg fits
  by a factor of $\sim 10$ we obtain a suitable match to the
  SN\,2001ig light-curves.
\label{fig:lt_curve_comparison}} 
\end{figure}

A comparison of the radio light-curves represented in
Table~\ref{tab:bumps} shows that the variations are generally unique
to each supernova.  However, as Figure~\ref{fig:lt_curve_comparison}
shows, the separation and intensity of the observed variations in
SN\,2003bg and the Type IIb SN\,2001ig are strikingly similar.  By
simply scaling our SN\,2003bg light-curve fits by a factor of $\sim 
0.1$ we obtain a reasonable match to the SN\,2001ig radio evolution.
With this scaling we derive the following physical parameters for the
SN\,2001ig radio emitting ejecta: $r\approx 4.2\times
10^{15}(t/t_0)^{0.8}$ cm, $\overline{v}\approx 0.16 c(t/t_0)^{-0.2} $,
$B\approx 3.8(t/t_0)^{-1}$ G and $E\approx 1.3\times 10^{48}
(t/t_0)^{0.4}$ erg for $\epsilon_e=\epsilon_B=0.1$.  The electron
number density is thus given by $n_e\approx 3.0\times 10^5
(r/r_0)^{-2}~\rm cm^{-3}$ with successive peaks of relative intensity
comparable to those observed in SN\,2003bg (Table~\ref{tab:peaks}) and
an associated mass loss rate of $\dot{M}\approx 8.6\times 10^{-5}~\rm
M_{\odot}~yr^{-1}$ ($v_w=1000~\rm km~s^{-1}$).  We note that this mass
loss estimate is comparable to that reported by \citet{rss+04}, but
they adopt free-free absorption for the dominant absorption process
and a progenitor wind velocity of $10~\rm km~s^{-1}$, typical of red
supergiant Type II SN progenitors.  Our SSA modeling, however, shows
the ejecta velocity of SN\,2001ig is consistent with those of SNe Ibc
and the mass loss rate is in fact comparable to Galactic Wolf-Rayet
stars, similar to the case of SN\,2003bg.

Figure~\ref{fig:lt_curve_comparison} shows that strong light-curve
deviations are observed for both SNe on timescales of $\sim 120$ and
300 days with evidence for a weaker feature at $t\sim 600$ days.
Making the reasonable approximation that the ejecta evolve as
$r\sim 5\times 10^{15} (t/t_0)^{0.8}$ cm, these variations correspond
to density enhancements at radial distances of $R\sim 4\times 10^{16}$
and $8\times 10^{16}$ cm.  Assuming a progenitor wind velocity of
$v_w=1000~\rm km~s^{-1}$, we infer ejection timescales of $\sim 12$
and 24 years before the SN explosion.  Moreover, as shown in the
Figure~\ref{fig:lt_curve_comparison} the strength of the light-curve
variations decreases with time implying that the largest enhancements
are closest to the explosion site.

In addition to their radio light-curves, the optical evolution for SNe
2003bg and 2001ig are also similar.  In both cases there was
spectroscopic evidence for a thin layer of hydrogen near the time of
optical maximum \citep{psk+01,hpt03}; however nebular spectra revealed
no hydrogen emission in either event (\citealt{fc02}; Hamuy, private
communication)\footnote{Given the similarity between the two SNe, it
is natural to question whether earlier observations of SN\,2001ig
would have revealed a Type Ic spectrum as was observed for SN\,2003bg
at $t\sim 6$ days past explosion.}.  This evolution is characteristic
of Type IIb supernovae (e.g. SN\,1993J; \citealt{fmb94}).  However,
their inferred ejecta velocities resemble more closely those of SNe
Ibc and therefore imply compact (Wolf-Rayet) progenitors.  We conclude
that these two events are intermediate between Types IIb and Ibc, and
show more overall similarities to hydrogen-poor events.  Moreover, the
likeness of their observed radio and optical evolution may indicate
similar environments and progenitor evolutions.

\section{The Progenitors of SNe 2003bg and 2001ig}
\label{sec:prog}

The overall resemblance of SN\,2003bg and SN\,2001ig as outlined in
the previous section suggests that a similar process is responsible
for the observed density variations.  Based on the location ($R\gtrsim
5\times 10^{16}$ cm) and relative intensity (factor of $\lesssim 2$)
of the density enhancements, we find that Cases 1 and 4 (a variable mass loss
rate and the tidal effects of a binary companion, respectively) provide 
the most likely explanations for the inferred circumstellar structure.

Noting the abrupt fluctuations in the SN\,2001ig radio light-curves,
\citet{rss+04} recently proposed a binary induced pinwheel nebula for
the inferred episodic density enhancements.  In this scenario, the
colliding stellar winds produce a smooth dusty plume that spirals in
the plane the binary orbit.  A highly eccentric orbit is required to
produce structure at well-defined intervals and a favorable viewing
angle is necessary to reproduce the observed flux variations.  Here
the location and intensity of the pinwheel structure is determined by
the binary parameters.  \citet{rms06} argue further support for this
model based on the discovery of a late-B - late-F supergiant
coincident (rms uncertainty of 0.03 arcsec, corresponding to 1.7
pc at $d\approx 11.5$ Mpc) with the SN\,2001ig optical position which
they interpret as the putative binary companion.  However, in cases of
observed pinwheel nebulae (e.g. \citealt{tmd99}), the dust plumes are
attributed to the strong wind-wind interaction between a Wolf-Rayet
and O- or early-B star companion.  Taken together with the substantial
uncertainty in the position of the putative companion, this suggests
that a binary induced pinwheel nebula for SN\,2001ig is not clearly
substantiated.

Adopting a pinwheel model for both SNe 2003bg and 2001ig would imply that
the geometry of their circumstellar nebulae are nearly identical and
thus they share remarkably similar binary systems.  Moreover, this
model would imply that our viewing angle with respect to their
circumstellar nebulae must also be very similar.  While this scenario
cannot be ruled out, it is clearly unlikely.

Assuming a single star progenitor model, the mass loss rates derived
through our SSA modeling for SNe 2003bg and 2001ig imply massive
Wolf-Rayet progenitors.  Broadband observations of Galactic WR stars
show that the circumstellar media for these events are often
significantly disturbed.  In particular, X-ray observations show
evidence for colliding stellar winds and dense clumps
(e.g. \citealt{h03}).  If the progenitors of SNe 2003bg and 2001ig
were single WR stars then the observed density variations could be
produced by a series of enhanced mass loss episodes.  To be consistent
with the radio data, the mass loss ejections must ramp up as the star
nears explosion.  In this scenario, the observed variation timescale
and the lack of CSM structure at radii inward of $R\sim 4\times
10^{16}$ cm may be explained by a quasi-periodic mechanism which
drives enhanced mass loss episodes every $\sim 12$ years.  If
accompanied by wind velocity variations, these episodes could produce
shells of circumstellar material, overall consistent with the
circularly symmetric density structures we infer for SNe 2003bg and
2001ig.  The observed similarity between SNe 2003bg and 2001ig implies
that these mass loss episodes may be common among some fraction of
hydrogen-poor events.

These clues suggest that episodic variations in the progenitor mass
loss rate and/or wind velocity provide a reasonable explanation to the
radio light-curve variations observed for both SNe 2003bg and 2001ig.
Given the predicted timescale for the mass ejections, these two SNe
provide some of the best evidence for unusual mass loss evolution
immediately preceding the supernova explosion.  While the presence of
a binary companion cannot be ruled out for either of these
hydrogen-poor events, additional studies of short timescale
variability in radio supernovae will enable discrimination between single
and binary progenitor models.

\section{Conclusions}
\label{sec:conc}
We report extensive radio and X-ray observations of SN\,2003bg.  The
spectroscopic evolution of this unusual supernova shows a transition
from a broad-lined Type Ic to a hydrogen-rich Type II and later to a
typical hydrogen-poor Type Ibc.  This evolution strengthens the
connection between Type II and Type Ibc (including broad-lined)
events.  We show that the extraordinarily luminous radio emission for
a SN Ibc is well described by a self-absorption dominated synchrotron
spectrum while the observed X-ray emission at $t\approx 30$ days is
adequately fit by inverse Compton scattering of the optical photons
off of the synchrotron emitting electrons.  Our radio model implies a
sub-relativistic ejecta velocity, $\overline{v}\approx 0.24 c$, and a
size of $r\approx 6.2\times 10^{15}$ cm at $t_0\approx 10$ days.  This
analysis emphasizes that broad optical absorption lines do not imply
relativistic ejecta.  We find that the total energy of the radio
emitting region evolves as $E\approx 7.3\times 10^{48} (t/t_0)^{0.4}$
erg assuming equipartition of energy between relativistic electrons
and magnetic fields ($\epsilon_e=\epsilon_B=0.1$).  The circumstellar
density is well described by stellar wind profile, $\propto r^{-2}$,
with modest (factor of $\sim 2$) episodic enhancements which produce
abrupt achromatic flux variations.  We show that free-free absorption
does not contribute significantly to the radio spectrum and estimate
an average mass loss rate of $\dot{M}\approx 3 \times 10^{-4}~\rm
M_{\odot}~yr^{-1}$, consistent with observed values for local
Wolf-Rayet stars.

Comparison with other events reveals that $\sim 50\%$ of radio
supernovae show similar short timescale light-curve variations which
are attributable to circumstellar density irregularities.  This
compilation emphasizes that abrupt radio light-curve variations cannot
be used as a reliable proxy for an engine-driven explosion, including
off-axis gamma-ray bursts.

Finally, the radio light-curves and spectroscopic evolution for
SN\,2003bg are strikingly similar to those of SN\,2001ig, suggestive
of a common progenitor evolution for these two events.  The overall
similarity of SNe 2003bg and 2001ig to radio SNe Ibc is suggestive of
a compact Wolf-Rayet progenitor model.  Based on the relative intensity of the
inferred density enhancements, we conclude that the progenitors of
hydrogen-poor SNe 2003bg and 2001ig experienced quasi-periodic mass
loss episodes just prior to the SN explosion.

\acknowledgments We are indebted to Barry Clark for his generous
scheduling of the VLA for this project.  We thank M. Hamuy and A.
Filippenko for making their optical SN\,2003bg data available to us
during this study.  The authors thank Re'em Sari, Doug Leonard, Edo
Berger, Michael Rupen, Paolo Mazzali, Avishay Gal-Yam, P. Brian
Cameron, Derek Fox, Ehud Nakar and Kurt Weiler for useful discussions.
Caltech SN research is supported by NSF and NASA grants.  AMS
acknowledges support by the National Radio Astronomy Observatory
Graduate Summer Student Research Assistantship program and the NASA
Graduate Student Research Program.  RAC acknowledges NSF grant
AST-0307366.

\bibliographystyle{apj1b}

\clearpage

\begin{deluxetable}{lrrrrrrrc}
\rotate
\tablecaption{VLA radio flux density measurements of SN\,2003bg}
\label{tab:vla}
\tablewidth{0pt}
\tablehead{ \colhead{Date} & \colhead{$\Delta t$} & \colhead{$F_{\nu,1.43\rm~GHz}$} & \colhead{$F_{\nu,4.86\rm~GHz}$} & \colhead{$F_{\nu, 8.46\rm~GHz}$} & \colhead{$F_{\nu,15.0\rm~GHz}$} & \colhead{$F_{\nu,22.5\rm~GHz}$} &  \colhead{$F_{\nu,43.3\rm~GHz}$} & \colhead{Array} \\
\colhead{(UT)} & \colhead{(days\tablenotemark{\dagger})} & \colhead{(mJy)} & \colhead{(mJy)} & \colhead{(mJy)} & \colhead{(mJy)} & \colhead{(mJy)} & \colhead{(mJy)} & \colhead{Config.} \\
}
\startdata
2003 Mar 4  & 10    & \nodata           & \nodata           & $2.51\pm 0.07$  & \nodata           & \nodata            & \nodata            & D \\
2003 Mar 6  & 12    & $0.42\pm 0.28$  & \nodata           & $3.86\pm 0.10$  & \nodata           & $30.77\pm 0.71$  & \nodata            & D \\
2003 Mar 17 & 23    & $0.55\pm 0.20$  & $2.75\pm 0.11$  & $12.19\pm 0.26$ & $47.03\pm 0.98$ & $106.30\pm 2.16$ & \nodata            & D \\
2003 Mar 29 & 35    & \nodata           & $8.87\pm 0.20$  & $24.72\pm 0.50$ & $62.11\pm 1.26$ & $85.39\pm 1.72$  & $64.07\pm 1.95$  & D \\
2003 Apr 11 & 48    & \nodata           & $14.88\pm 0.32$ & $40.34\pm 0.81$ & $71.51\pm 1.46$ & $74.58\pm 1.53$  & $31.75\pm 1.43$  & D \\
2003 Apr 21 & 58    & \nodata           & $22.37\pm 0.46$ & $51.72\pm 1.04$ & $69.31\pm 1.41$ & $39.91\pm 0.89$  & $20.56\pm 5.87$  & D \\
2003 Apr 26 & 63    & \nodata           & $24.62\pm 0.50$ & $49.64\pm 1.00$ & $64.75\pm 1.32$ & $32.48\pm 0.71$  & $15.62\pm 2.53$  & D \\
2003 May 6  & 73    & \nodata           & $28.85\pm 0.59$ & $46.20\pm 0.93$ & $41.18\pm 0.87$ & $30.53\pm 0.67$  & $10.18\pm 2.55$  & D \\
2003 May 18 & 85    & \nodata           & $31.14\pm 0.68$ & $38.648\pm 0.79$ & $19.39\pm 0.54$ & $9.63\pm 1.13$   & \nodata            & DnA \\
2003 May 24 & 91    & \nodata           & $32.42\pm 0.69$ & $33.85\pm 0.71$ & $17.34\pm 0.83$ & $10.70\pm 2.10$  & \nodata            & DnA \\
2003 Jun 17 & 115   & \nodata           & $39.07\pm 0.79$ & $45.74\pm 0.92$ & $32.14\pm 0.68$ & $21.07\pm 0.44$  & \nodata            & A \\
2003 Jun 31 & 129   & $9.27\pm 0.30$  & $47.37\pm 0.95$ & $53.94\pm 1.08$ & $41.02\pm 0.84$ & $30.19\pm 1.70$  & \nodata            & A \\
2003 Jul 4  & 132   & $8.77\pm 0.36$  & $48.81\pm 0.98$ & $54.27\pm 1.09$ & $40.92\pm 0.83$ & \nodata            & \nodata            & A \\    
2003 Jul 14 & 142   & $10.10\pm 0.48$ & $52.69\pm 1.06$ & $54.83\pm 1.10$ & $39.24\pm 0.81$ & \nodata            & \nodata            & A \\
2003 Jul 29 & 157   & $11.27\pm 0.39$ & $52.81\pm 1.06$ & $48.43\pm 0.97$ & $30.88\pm 0.63$ & \nodata            & \nodata            & A \\
2003 Aug 2  & 161   & $11.38\pm 0.44$ & $50.84\pm 1.02$ & $47.43\pm 0.95$ & $29.17\pm 0.61$ & \nodata            & \nodata            & A \\
2003 Aug 22 & 181   & $12.79\pm 0.35$ & $45.19\pm 0.91$ & $35.76\pm 0.72$ & $21.06\pm 0.48$ & $11.00\pm 0.61$  & \nodata       & A \\
2003 Sep 11 & 201   & $14.07\pm 0.37$ & $40.10\pm 0.81$ & $31.35\pm 0.63$ & $16.58\pm 0.44$ & $6.59\pm 0.43$   & \nodata   & BnA \\
2003 Sep 24 & 214   & $17.07\pm 0.40$ & $39.65\pm 0.80$ & $28.67\pm 0.58$ & $16.54\pm 0.39$ & $10.13\pm 0.48$  & \nodata            & BnA \\
2003 Oct 7  & 227   & $15.94\pm 0.34$ & $36.45\pm 0.73$ & $27.38\pm 0.55$ & $15.58\pm 0.35$ & $7.84\pm 0.47$   & \nodata            & BnA \\
2003 Oct 22 & 242   & $19.31\pm 0.49$ & $36.53\pm 0.74$ & $24.57\pm 0.50$ & \nodata           & $9.38\pm 0.23$   & \nodata            & B \\
2003 Nov 4  & 255   & $19.23\pm 0.48$ & $34.56\pm 0.70$ & $22.30\pm 0.46$ & $12.71\pm 0.35$ & $7.86\pm 0.31$   & \nodata            & B \\
2003 Nov 15 & 266   & $23.05\pm 0.59$ & $33.37\pm 0.68$ & $21.67\pm 0.44$ & $11.31\pm 0.30$ & $7.79\pm 0.25$   & \nodata            & B \\
2003 Dec 4  & 285   & $22.99\pm 0.54$ & $33.12\pm 0.67$ & $21.31\pm 0.43$ & $12.02\pm 0.28$ & $7.74\pm 0.19$   & \nodata            & B \\
2003 Dec 19 & 300   & $24.13\pm 0.54$ & $33.19\pm 0.67$ & $20.88\pm 0.42$ & $11.67\pm 0.28$ & $7.76\pm 0.19$   & \nodata            & B \\
2004 Jan 14 & 326   & $24.50\pm 0.59$ & $32.90\pm 0.66$ & $20.33\pm 0.41$ & $11.36\pm 0.28$ & $6.84\pm 0.19$   & \nodata            & B \\
2004 Jan 25 & 337   & $28.30\pm 0.61$ & $31.66\pm 0.64$ & $19.85\pm 0.40$ & $9.08\pm 0.26$  & $5.96\pm 0.18$   & \nodata            & CnB \\
2004 Feb 8  & 351   & $23.88\pm 0.62$ & $30.90\pm 0.63$ & $18.84\pm 0.38$ & $9.65\pm 0.26$  & $6.72\pm 0.16$   & $3.36\pm 0.57$   & CnB \\
2004 Feb 25 & 368   & $31.99\pm 0.76$ & $28.73\pm 0.58$ & $17.14\pm 0.35$ & $9.72\pm 0.25$  & $5.72\pm 0.17$   & \nodata            & C \\
2004 Apr 2  & 405   & \nodata           & \nodata           & $14.61\pm 0.30$ & \nodata           & $7.48\pm 0.39$   & \nodata            & C \\
2004 Apr 7  & 410   & $33.27\pm 0.72$ & $23.88\pm 0.50$ & $14.49\pm 0.31$ & \nodata           & \nodata            & \nodata            & C \\
2004 Apr 21 & 424   & $27.93\pm 0.63$ & $22.96\pm 0.47$ & $14.16\pm 0.29$ & $6.41\pm 0.25$  & $2.73\pm 0.13$   & $0\pm 0.39$       & C \\
2004 Apr 31 & 434   & $24.82\pm 0.57$ & $20.98\pm 0.43$ & $13.25\pm 0.27$ & $5.75\pm 0.21$  & $3.05\pm 0.24$   & \nodata            & C \\
2004 May 2  & 435   & \nodata           & \nodata           & $13.08\pm 0.27$ & \nodata           & \nodata            & $0.38\pm 0.23$   & C \\
2004 Jun 29 & 493   & $26.60\pm 0.76$ & $16.20\pm 0.34$ & $10.04\pm 0.21$ & $6.59\pm 0.36$  & $3.78\pm 0.51$   & $1.31\pm 0.38$   & D \\
2004 Aug 8  & 533   & $23.65\pm 0.74$ & $14.60\pm 0.33$ & $8.92\pm 0.22$  & $5.72\pm 0.40$  & $3.25\pm 0.58$   & \nodata            & D \\
2004 Nov 15 & 632   & \nodata           & \nodata           & $6.23\pm 0.41$  & $4.47\pm 1.08$  & $2.74\pm 0.57$   & \nodata            & A \\
2004 Nov 29 & 646   & $23.70\pm 0.51$ & $10.82\pm 0.23$ & \nodata           & \nodata           & \nodata            & \nodata            & A \\
2005 Jan 24 & 702   & $23.23\pm 0.603$ & $10.21\pm 0.23$ & $6.18\pm 0.14$  & $3.43\pm 0.19$  & \nodata            & \nodata            & BnA \\
2005 Jan 31 & 709   & \nodata           & \nodata           & \nodata           & \nodata           & $1.86\pm 0.10$   & \nodata            & BnA \\
2005 Mar 19 & 756   & \nodata           & \nodata           & $4.62\pm 0.16$  & $2.52\pm 0.21$  & \nodata            & \nodata            & B \\
2005 Apr 11 & 779   & $23.15\pm 0.78$ & $8.49\pm 0.21$  & \nodata           & \nodata           & \nodata            & \nodata            & B \\
2005 May 22 & 820   & $21.43\pm 0.93$ & $7.56\pm 0.21$  & $3.93\pm 0.13$  & $2.69\pm 0.85$  & $0.85\pm 0.23$   & \nodata            & B \\
2005 Aug 12 & 902   & $21.71\pm 1.42$ & $8.23\pm 0.20$  & $4.69\pm 0.12$  & $2.56\pm 0.23$  & $2.22\pm 0.35$   & $1.31\pm 0.32$   & C \\
2005 Oct 27 & 978   & $21.84\pm 0.29$ & $7.83\pm 0.08$  & $4.48\pm 0.07$  & \nodata & \nodata & \nodata & DnC \\ 
\enddata
\tablenotetext{\dagger}{Days since explosion assuming an explosion date of 2003 February 22 UT.}
\label{tab:vla}
\end{deluxetable}

\clearpage

\begin{deluxetable}{lrlrrrrrr}
\rotate
\tablecaption{Spectral models for {\it Chandra} ACIS-S Observations of SN\,2003bg}
\tablewidth{0pt}
\tablehead{ 
\colhead{Date} & \colhead{$\Delta t$} & \colhead{Model} & \colhead{$N_{\rm H}$\tablenotemark{\ddagger}} & \colhead{$\Gamma$\tablenotemark{\ddagger}} & \colhead{$kT$\tablenotemark{\ddagger}} & \colhead{$\chi^2/dof$} & \colhead{Flux (0.3-10 keV)\tablenotemark{\ddagger}} & \colhead{$\beta_{RX}$\tablenotemark{*}}
\\
\colhead{(UT)}& \colhead{(days)} & \colhead{} & \colhead{($10^{22}~\rm cm^{-2}$)} & \colhead{} & \colhead{(keV)} & \colhead{} & \colhead{(erg~cm$^{-2}~\rm s^{-1}$)} & \colhead{}
}
\startdata
2003 March 24.5 & 30 & Power Law & 0.013\tablenotemark{\dagger} & $1.7\pm 0.2$ & \nodata & 24/20 & $(9.3\pm 0.4)\times 10^{-14}$ & -0.94 \\
\nodata & \nodata & \nodata & $0.054\pm 0.033$ & $1.9\pm 0.3$ & \nodata & 24/19 & $(9.6\pm 0.4)\times 10^{-14}$ & -0.94 \\
\nodata & \nodata & Th. Brem. & 0.013\tablenotemark{\dagger} & \nodata & $4.3\pm 0.9$ & 26/20 & $(7.8\pm 0.9)\times 10^{-14}$  & -0.95 \\
\nodata & \nodata & \nodata & $< 0.018$ & \nodata & $4.5\pm 0.9$ & 25/19 & $(7.7\pm 0.9)\times 10^{-14}$ & -0.95 \\
\hline
2003 June 22.3 & 120 & Power Law & 0.013\tablenotemark{\dagger} & $1.7$ & \nodata & 5/7 & $(1.2\pm 0.4)\times 10^{-14}$ & -0.89 \\
\nodata & \nodata & Th. Brem. & 0.013\tablenotemark{\dagger} & \nodata & $4.4\pm 1.6$ & 4/7 & $(1.2\pm 0.2)\times 10^{-14}$ & -0.97 \\
\enddata
\tablenotetext{\ddagger}{All errors represent 90\% confidence levels.}
\tablenotetext{\dagger}{$N_{\rm H}$ fixed to the Galactic value (see \S\ref{sec:cxo}).}
\tablenotetext{*}{Spectral index between optically thin radio and X-ray flux density.}
\label{tab:cxo}
\end{deluxetable}

\clearpage

\begin{deluxetable}{lrrrrrrrr}
\rotate
\tablecaption{Radio Light-Curve Peaks for SN\,2003bg}
\tablewidth{0pt}
\tablehead{ \colhead{Phase} & \colhead{$\Delta t$} & \colhead{$r$} & \colhead{$C_f$\tablenotemark{\dagger}} & \colhead{$C_{\tau}$\tablenotemark{\dagger}} & \colhead{$\nu_m$\tablenotemark{\dagger}} & \colhead{$n_e$\tablenotemark{\ddagger}} & \colhead{$B$\tablenotemark{\ddagger}} & \colhead{$E$\tablenotemark{\ddagger}} \\
\colhead{} & \colhead{(days)} & \colhead{($\times 10^{16}$ cm)} & \colhead{($\times 10^{-51}~\rm g~s^{1/2}$)} & \colhead{($\times 10^{38}~\rm s^{-3.6}$)} & \colhead{(GHz)} & \colhead{(jump)} & \colhead{(jump)} & \colhead{(jump)}
}
\startdata
1 & $\lesssim 110$ & $\lesssim 6.2$ & 3.1 & 6.2 & 0.11 & \nodata & \nodata & \nodata \\ 
2 & 120-250 & 6.7-12 & 2.7 & 23 & 0.15 & 1.8 & 1.3 & 1.8 \\
3 & 300-500 & 14-21 & 2.5 & 46 & 0.18 & 1.4 & 1.2 & 1.4 \\
4 & $\gtrsim 600$ & $\gtrsim 24$ & 2.4 & 67 & 0.19 & 1.2 & 1.1 & 1.2 \\
\enddata
\tablenotetext{\dagger}{Values computed at the reference time of $t_0=10$ days.}
\tablenotetext{\ddagger}{Factor by which the parameter is enhanced with respect to an extrapolation of the parameters from the previous phase.}
\label{tab:peaks}
\end{deluxetable}

\clearpage
\begin{deluxetable}{lcrrrr}
\tablecaption{Radio Supernovae with Strong Light-Curve Variations}
\tablewidth{0pt}
\tablehead{ \colhead{SN} & \colhead{Type} & \colhead{$t_{\rm peak}$} & \colhead{$\nu_{\rm peak}$} & \colhead{Variation} & \colhead{Ref.\tablenotemark{\dagger}} \\
\colhead{} & \colhead{} & \colhead{(years)} & \colhead{($\nu$ GHz)} & \colhead{Type} & \colhead{} 
}
\startdata
1957D & II & $<27$ & 4.9 & Abrupt Steepening & 1,2,3\\
1978K & IIn & 2.6  & 4.9 & Abrupt Steepening & 4 \\
1979C & IIL & 1.2 & 5.0 & Episodic Bumps & 1,5,6\\
1980K & IIL & 0.4 & 5.0 & Abrupt Steepening & 1,7\\ 
1986J & IIn & 3.8 & 4.9 & Abrupt Steepening & 8\\
1987A & II & 0.0093 & 0.84 & Abrupt Rise &  9,10\\
1988Z & IIn & 2.5 & 4.9 & Abrupt Steepening &  11 \\
1998bw & Ic & 0.030 & 4.9 & Episodic Bumps & 12,13,14 \\
2001em & Ic/II & $<2.5$ & 4.9 & Abrupt Rise & 15,16\\
2001ig & IIb & 0.20 & 5.0 & Episodic Bumps & 17 \\
2003bg & Ic/II & 0.27 & 4.9 & Episodic Bumps & 18 \\
2004C  & Ic & $<0.24$ & 4.9 & Abrupt Rise & 19 \\
2004cc & Ic & 0.068 & 8.5 & Abrupt Steepening & 19 \\
\enddata
\tablenotetext{\dagger}{References: 1 - \citet{wsp+86}; 2 - \citet{crb94}, 3 - \citet{smc+05}; 4 - \citet{srs+99}; 5 - \citet{wvp+92}; 6 - \citet{mwv+00}; 7 - \citet{mvw+98}; 8 - \citet{bbr02}; 9 - \citet{bcc+95}; 10 - \citet{c98}; 11 - \citet{wpv+02}; 12 - \citet{kfw+98}; 13 - \citet{lc99}; 14 - \citet{wpm01}; 15 - \citet{svs+04}; 16 - \citet{cc05}; 17 - \citet{rss+04}; 18 - this paper; 19 - Soderberg {\it et al.}, in prep.}
\label{tab:bumps}
\end{deluxetable}

\clearpage

\clearpage

\clearpage

\clearpage

\clearpage

\clearpage

\clearpage

\begin{figure}
\plotone{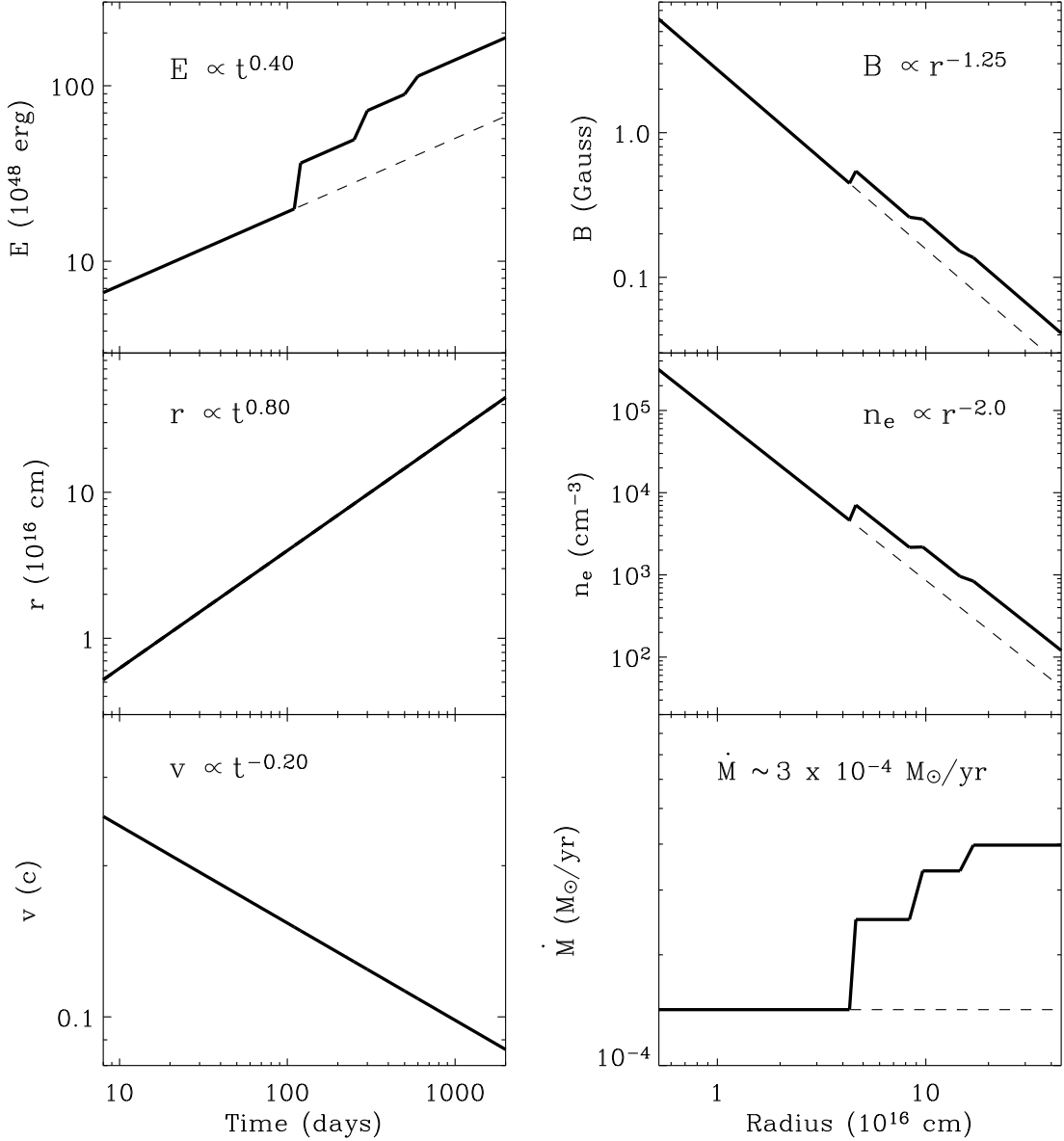}
\bigskip
\bigskip
\caption{Physical parameters for SN\,2003bg based on our synchrotron
  self-absorption model described in \S\ref{sec:model} and applied in
  \S\ref{sec:ssa}.  Left column: the temporal evolution from $t\approx
  10$ to 2000 days is shown for the energy of the emitting region,
  shock radius and average velocity (top to bottom).  Right column:
  the radial profile of the magnetic field, electron number density
  and mass loss rate are shown from $r\approx 5\times 10^{15}$ to
  $4\times 10^{17}$ cm.  An extrapolation of our SSA model with
  no density variations is shown for comparison (dashed line).
\label{fig:params}}
\end{figure}

\clearpage

\end{document}